\documentclass[a4paper,fleqn]{cas-sc}

\usepackage{graphicx}      
\usepackage[authoryear]{natbib}
\def\tsc#1{\csdef{#1}{\textsc{\lowercase{#1}}\xspace}}
\tsc{WGM}
\tsc{QE}

\usepackage{indentfirst}
\usepackage{array,multirow}
\usepackage{fancyhdr}
\usepackage{float}
\usepackage{booktabs}
\usepackage{here}
\usepackage{comment}
\usepackage{siunitx}
\usepackage{hyperref}
\usepackage{subcaption}
\usepackage{amsmath}
\usepackage{enumitem}
\usepackage{setspace}
\usepackage{xcolor}
\usepackage{cite}
\renewcommand{\thefootnote}{\roman{footnote}} 
\usepackage[lcgreekalpha]{stix2}

\usepackage{multirow}
\usepackage{here}
\usepackage{stfloats}
\usepackage{CJKutf8}
\usepackage{hyperref}
\usepackage{color}
\usepackage[right]{lineno}

\hypersetup{backref=true,       
    pagebackref=true,               
    hyperindex=true,                
    colorlinks=true,                
    breaklinks=true,                
    urlcolor= cyan,                
    linkcolor= cyan,                
    bookmarks=true,                 
    bookmarksopen=false,
    filecolor=cyan,
    citecolor=cyan,
    linkbordercolor=cyan
}

\newcommand{\etal}{et~al.\ }
\newcommand{\eg}{e.\,g.,\ }
\newcommand{\ie}{i.\,e.,\ }

\newenvironment{subsubfigure}[2][]{%
  \begin{subfigure}[#1]{#2}%
    \stepcounter{subsubfigure}%
}{%
    \addtocounter{subfigure}{-1}%
  \end{subfigure}%
}

\newcounter{subsubfigure}

\begin{document}

\let\WriteBookmarks\relax
\def\floatpagepagefraction{1}
\def\textpagefraction{.001}

\shorttitle{Path Indication and Motion Sickness in APMVs} 

\shortauthors{Ide \etal}  

\title[mode=title]{Driving Path Indication Reduces Motion Sickness and Influences Head Motion of Passengers in Autonomous Personal Mobility Vehicle}

\author[1]{Yuya Ide}

\author[1]{Hailong Liu}[orcid=0000-0003-2195-3380]
\cormark[1]
\cortext[1]{Corresponding author}
\ead{liu.hailong@is.naist.jp}

\author[1]{Takahiro Wada}[orcid=0000-0002-4518-8903]

\affiliation[1]{organization={Nara Institute of Science and Technology (NAIST)},
    addressline={8916-5 Takayama-cho}, 
    city={Ikoma},
    postcode={630-0192}, 
    state={Nara},
    country={Japan}}

\begin{abstract}
Autonomous personal mobility vehicles (APMVs) are novel smart mobility devices designed to provide automated individual transportation in indoor or mixed-traffic environments.
However, in such environments, frequent pedestrian avoidance maneuvers may cause rapid steering adjustments and passive postural responses from passengers, thereby increasing the risk of motion sickness.
This study investigated whether indicating the future driving path could mitigate motion sickness in APMV passengers. 
A mixed-design experiment was conducted with 40 participants under two self-reported genders as a between-subject factor (\ie male and female), two driving paths as a between-subject factor (\ie irregular and regular) and three driving conditions as a within-subject factor (\ie manual driving (MD), automated driving without path indication (AD w/o path), and automated driving with path indication (AD w/ path)).
Motion sickness was evaluated using the Motion Illness Symptom Classification (MISC), and head motion was assessed by calculating the delay time of participants' head yaw rate relative to APMV's yaw rate in the turning direction. 

The results showed that driving condition was the only factor that significantly affected both motion sickness and head-motion delay. 
Compared with the AD w/o path condition, both the MD and AD w/ path conditions were associated with lower motion sickness severity, longer motion sickness onset latency, and earlier head motion relative to vehicle motion. 
Notably, the AD w/ path condition achieved motion sickness levels comparable to those in the MD condition. Furthermore, repeated-measures correlation analysis showed significant associations between head-motion delay and all MISC metrics but the underlying physiological mechanism remains to be elucidated.

These findings suggest that presenting information about future driving path can mitigate motion sickness in APMV passengers. 
Moreover, the temporal characteristic of head motion relative to vehicle's motion may be considered a potential factor associated with motion sickness in autonomous mobility.
\end{abstract}


\maketitle

\section{INTRODUCTION}
\subsection{Background}

Autonomous personal mobility vehicles (APMVs) are novel smart mobility devices designed to provide automated individual transportation in indoor or mixed-traffic environments (\eg department stores, city sidewalks)~\citep{harada2024pedestrian,omori2024autonomous}.
APMVs are designed for a wide range of users, including not only older adults and people with mobility challenges, but also anyone seeking convenient short-distance transportation~\citep{liu2022implicit,liu2024silent}.
These APMVs are typically equipped with SAE Level 3 to 5 automated driving systems (ADS), enabling them to navigate in pedestrian-rich environments with minimal or no user intervention.

In such driving environments with high pedestrian density, APMVs are required to frequently adjust their trajectories in response to dynamic and unpredictable pedestrian movement~\citep{pham2015evaluation,isono2022autonomous}. 
Thus, passengers on the APMVs, having little controllability or predictability of the vehicle's driving behavior, often become passive recipients of the motion and the risk of motion sickness increase, potentially resulting in severe discomfort~\citep{liu2024subjective}.
Therefore, reducing motion sickness is an important task in improving the riding comfort of APMV passengers.

\subsection{Related Works}

Motion sickness causes symptoms such as dizziness, nausea, and headache~\citep{MISC}, and susceptibility to these symptoms varies across individuals~\citep{money1970motion, golding2006motion}.
Several theories have been proposed to explain the mechanisms of motion sickness. 
The postural instability theory argues that motion sickness arises when effective strategies for maintaining postural stability are lacking, particularly in unfamiliar environments \citep{riccio1991ecological}. Similarly, neural mismatch theory explains motion sickness as a consequence of a mismatch between sensory inputs and expected motion patterns represented in the internal model formed through prior experience~\citep{reason1978motion2}.
Based on this theory, \citet{bles1998motion} proposed the subjective vertical conflict theory, which attributes motion sickness to a conflict between the vertical direction sensed by the sensory organs and that estimated by the central nervous system. 
Mathematical models based on SVC theory have been widely used to predict and interpret motion sickness in emerging transportation systems, including automated vehicles (AVs) \citep{ukita2020simulation,buchheit2021motion} and APMVs \citep{liu2024subjective}.

In transportation domain, motion sickness may become more severe at intersections and sharp curves in urban areas because it is known that humans are particularly susceptible to motion sickness from low-frequency horizontal acceleration in the range around 0.2 Hz~\citep{turner1999motion,golding2001motion,beard2013discomfort}.
Moreover, it is said that passengers in autonomous vehicles face a higher risk of experiencing motion sickness than in human-operated vehicles~\citep{diels2016self}.
For this issue, improving passengers' ability to anticipate vehicle motion has been shown to be a promising approach for mitigating motion sickness. 
For example, in an automated driving context, \citet{karjanto2018effect} reported that presenting a light cue three seconds before a curve reduced motion sickness symptoms.
Similarly, \citet{kuiper2020knowing} and \citet{reuten2024mitigating} showed in simulator-based studies that anticipatory cues delivered through auditory or haptic modalities can help passengers predict upcoming vehicle motions and reduce motion sickness.
However, the relationship between predictability of vehicle motions and motion sickness is not always straightforward. 
For example, \citet{wijlens2025road} reported no significant difference in passenger motion sickness between repetitive and non-repetitive turning patterns in automated driving, despite the general expectation that more predictable motion should reduce motion sickness.

Evidence from manually driven vehicles also suggests that anticipation of vehicle motion is related to motion sickness~\citep{reuten2024anticipatory}. 
Drivers are generally less susceptible to motion sickness than passengers \citep{rolnick1991driver,diels2016self,wada2018analysis}.
In addition, drivers and passengers show different head motions during turning. 
\citet{zikovitz1999head} reported that drivers tend to actively tilt their torso and head toward the turn direction, whereas passengers show a more passive tilt in the opposite direction. 
\citet{Fujisawa20096} further reported a close relationship between vehicle lateral acceleration and drivers' head roll angle, indicating that drivers actively adjust their head posture in response to vehicle motion. 
Consistent with this view, \citet{wada2012can} showed that when passengers actively imitated a driver-like head tilt strategy toward the centripetal direction, their motion sickness symptoms were significantly reduced. 
These findings suggest that more anticipatory and driver-like responses to vehicle motion may be associated with reduced motion sickness.

Visual access to the upcoming road environment has also been suggested to contribute to motion prediction~\citet{kuiper2018looking}.
Some studies have reported that blocking passengers' forward view in vehicles increases motion sickness~\citep{griffin2004visual,tatsuno2024considerations}, possibly because passengers can use road geometry to anticipate future vehicle behavior. 
This could be considered that passengers can use road geometry to anticipate future vehicle behavior, such as that a curve ahead suggests the vehicle will turn.
From this perspective, presenting upcoming driving-path information may improve passengers' ability to anticipate vehicle motion and thereby mitigate motion sickness. This issue may be especially important in APMVs. 
Unlike conventional cars, APMVs are often used in shared spaces and indoor environments where clearly defined lane markings are absent, potentially making it more difficult for passengers to anticipate the future path of the vehicle. 
Accordingly, the lack of explicit path information in such environments may further reduce motion predictability and increase passengers’ susceptibility to motion sickness.

\subsection{Purpose}

Based on the aforementioned related studies, this study aims to investigate whether presenting information about the APMV's driving path is associated with passengers' motion sickness progression and head movement patterns during the ride. 
Additionally, passenger gender and driving path (\ie irregular and regular paths) are also included as factors to assess their associations with these outcomes.

\section{Method}

This study employed a $2 \times 2 \times 3$ mixed factorial design to examine whether presenting information about the APMV's upcoming movements is associated with passengers' motion sickness progression and head movement patterns. 
Specifically, two self-reported gender categories, \ie male and female, and two driving paths, \ie irregular and regular paths, were treated as between-subject factors, whereas three driving conditions, \ie manual driving (MD), automated driving without path indication (AD w/o path), and automated driving with path indication (AD w/ path), were treated as a within-subject factor. 
Detailed descriptions of these factors are provided in the following sections.

This study was carried out with the approval of the Research Ethics Committee of Nara Institute of Science and Technology
(No. 2021-I-38-3).

\subsection{Equipments}

\begin{figure}[!h]
    \centering
    \includegraphics[width=0.6\linewidth]{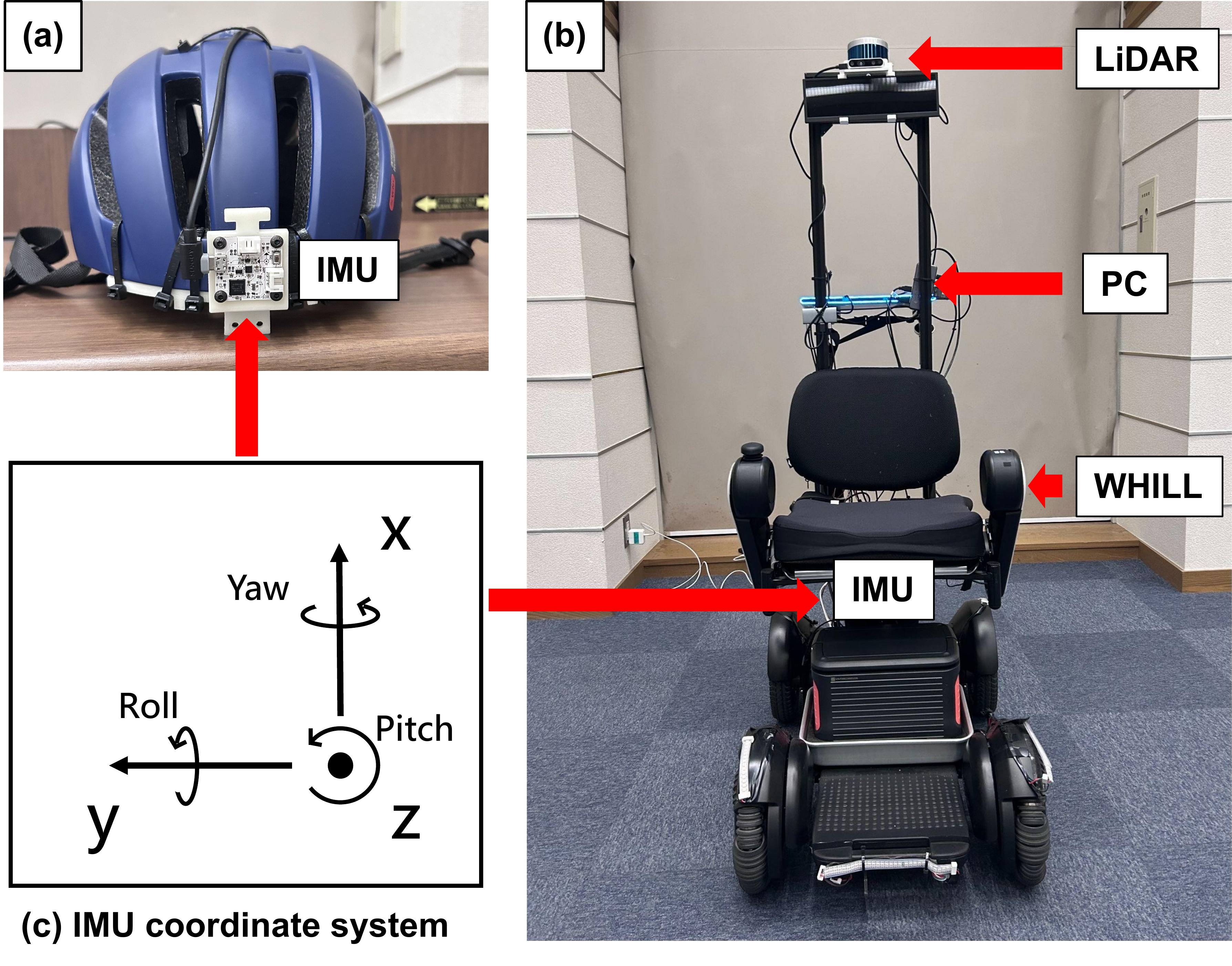}
\caption{Equipment used in the experiment. (a) Helmet with IMU to measure angular velocity of the head. (b) APMV in which the participants in the experiment enviroment. (c) IMU coordinate system attached to the helmet and APMV.}
    \label{fig:Appra}
    \vspace{2mm}
\end{figure}

An APMV with an SAE level 4 automated driving system based on the ROS Noetic used in the experiment consisted of a robotic wheelchair (WHILL CR) equipped with a LiDAR (Velodyne VLP-16) and an IMU (RT-USB-9axisIMU2), as shown in Fig.~\ref{fig:Appra}.
This APMV was autonomously controlled in real-time along a pre-recorded driving path.
Its maximum linear velocity was set to $6~\mathrm{km/h}$, and the maximum linear acceleration was limited to $1.7~\mathrm{m/s}^2$.

A helmet equipped with the same type of IMU as the one used on the APMV was prepared to measure the linear acceleration and angular velocity experienced by the passenger's head.
To ensure consistency and comparability of motion data, the IMU mounted on the helmet was oriented in the same direction as the IMU installed on the APMV.
These two IMUs recorded acceleration and angular velocity on three axes at a sampling rate of 100 Hz.

\subsection{Driving Path}

To examine the association between motion predictability of the APMV and passenger motion sickness, two driving paths were implemented in the experiment.
One is an irregular path with an unstructured trajectory, and another one is a regular path following a slalom trajectory.

\subsubsection{Irregular Path}

A 11.3 m $\times$ 12.0 m room was used as the experimental site for the irregular path, with uniform lighting employed to maintain consistent ambient illumination across all trials. 
To prevent passengers from anticipating APMV motions based on short-term repeated motion patterns, the driving path was intentionally designed to be unstructured and less predictable (see Fig.~\ref{fig:Irre_path}).

\subsubsection{Regular Path}
\begin{figure}[!ht] 
  \centering
   \begin{subfigure}{0.49\textwidth}
    \centering
    \includegraphics[width=1\linewidth]{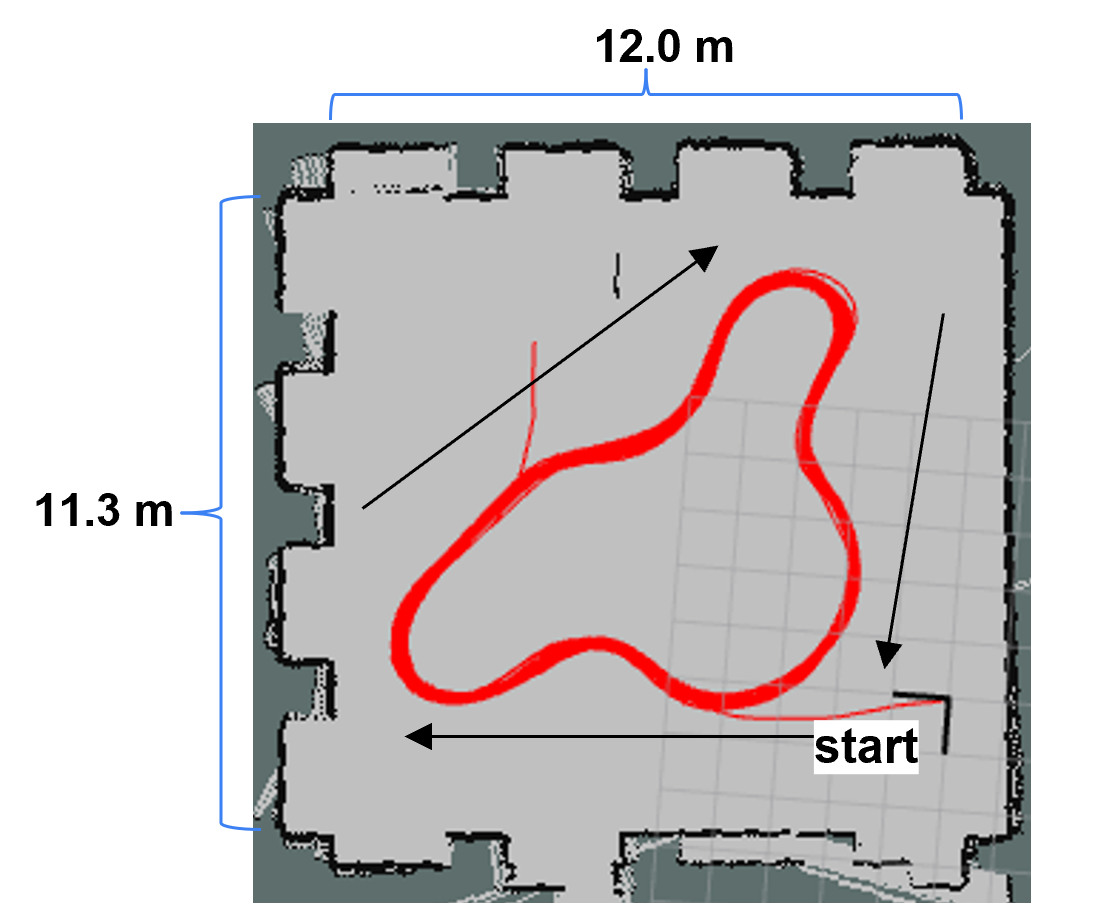}
    \caption{Irregular path}
    \label{fig:Irre_path}
  \end{subfigure}
  \begin{subfigure}{0.49\textwidth}
    \centering
    \includegraphics[width=0.69\linewidth]{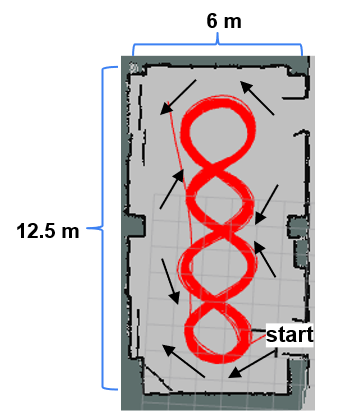}
    \caption{Regular path}
    \label{fig:reg_path}
  \end{subfigure}
\caption{Driving paths used in the experiment. (a) Irregular path with an unstructured trajectory. (b) Regular path following a slalom trajectory adapted from \citep{liu2024subjective}.}
  \label{fig:path}
\end{figure}

A 12.5 m $\times$ 6 m room was used as the experimental site for the regular path, with uniform lighting employed to maintain consistent ambient illumination across all trials. 
In contrast to the irregular condition, the driving path in the regular condition followed a short-term repeated motion pattern, \ie structured slalom pattern, allowing passengers to anticipate upcoming APMV motions (see Fig.~\ref{fig:reg_path}). 
This slalom driving path was implemented based on the experimental design of \citep{liu2024subjective}, in which the trajectory was shown to induce motion sickness symptoms in APMV passengers.

\vspace{2mm}

\begin{figure}[ht]
\centering

\begin{subfigure}{\linewidth}
  \centering
  \begin{subsubfigure}[b]{0.49\linewidth}
    \centering
    \includegraphics[width=\linewidth]{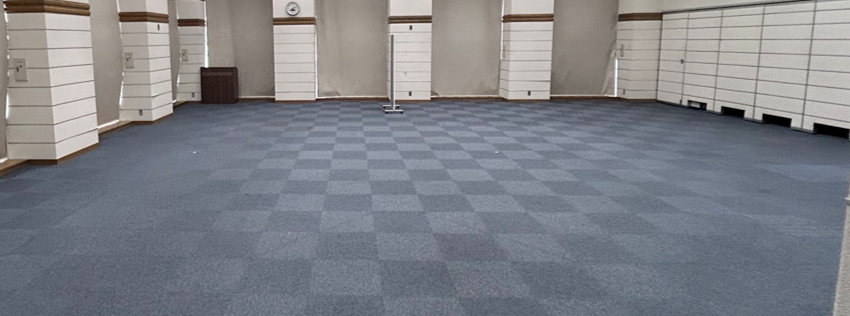}
    \caption{Irregular path}
    \label{AD w/o path Ire}
  \end{subsubfigure}
  \hfill
  \begin{subsubfigure}[b]{0.49\linewidth}
    \centering
    \includegraphics[width=\linewidth]{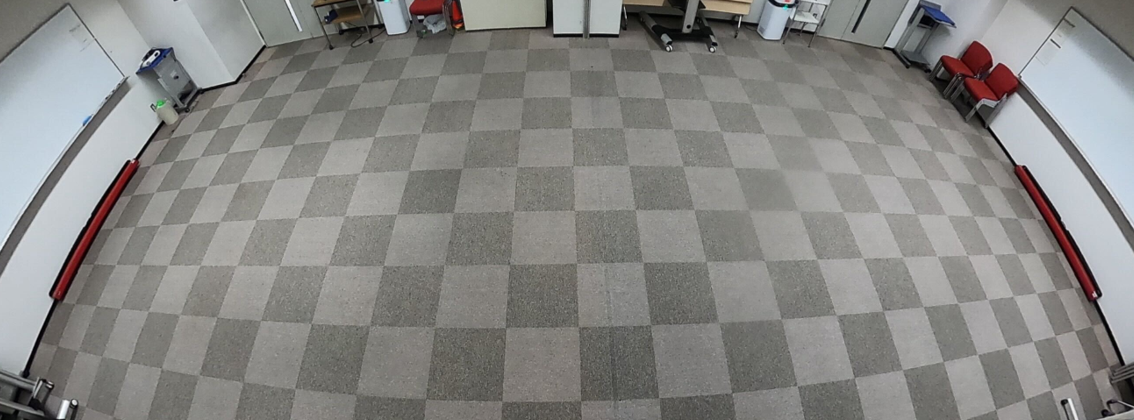}
    \caption{Regular path}
    \label{AD w/o path Re}
  \end{subsubfigure}
  \caption{Autonomous Driving without path (AD w/o path)}
  \label{fig:AD w/o path}
\end{subfigure}

\medskip

\begin{subfigure}{\linewidth}
  \centering
  \begin{subsubfigure}[b]{0.49\linewidth}
    \centering
    \includegraphics[width=\linewidth]{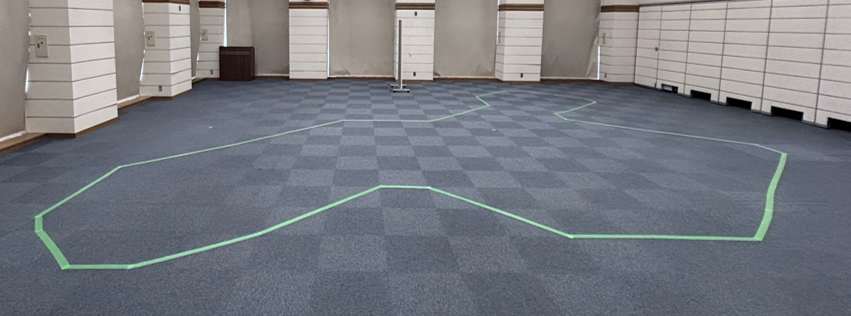}
    \caption{Irregular path}
    \label{AD w/ path Ire}
  \end{subsubfigure}
  \hfill
  \begin{subsubfigure}[b]{0.49\linewidth}
    \centering
    \includegraphics[width=\linewidth]{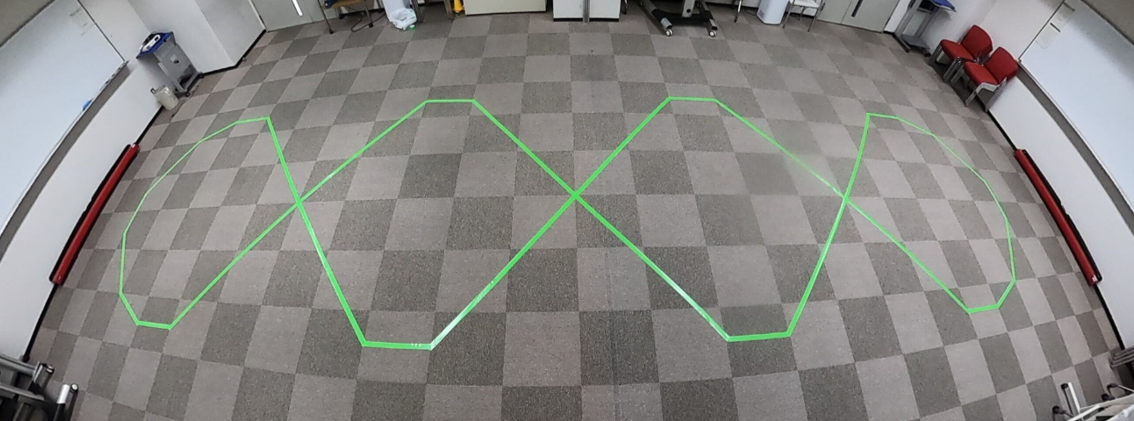}
    \caption{Regular path}
    \label{AD w/ path Re}
  \end{subsubfigure}
  \caption{Autonomous Driving with path (AD w/ path)}
  \label{fig:AD w/ path}
\end{subfigure}

\medskip

\begin{subfigure}{\linewidth}
  \centering
  \begin{subsubfigure}[b]{0.49\linewidth}
    \centering
    \includegraphics[width=\linewidth]{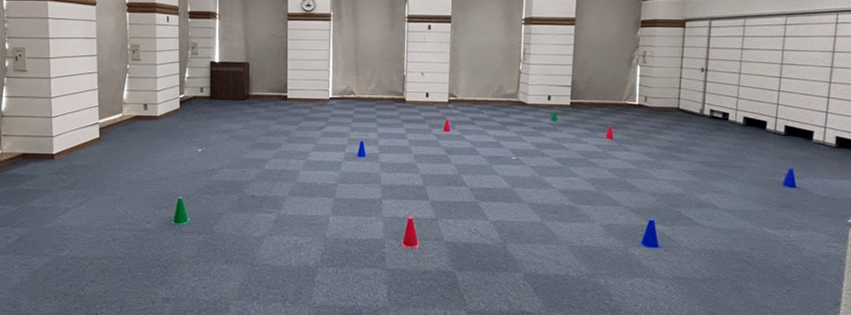}
    \caption{Irregular path}
    \label{MD_Ire}
  \end{subsubfigure}
  \hfill
  \begin{subsubfigure}[b]{0.49\linewidth}
    \centering
    \includegraphics[width=\linewidth]{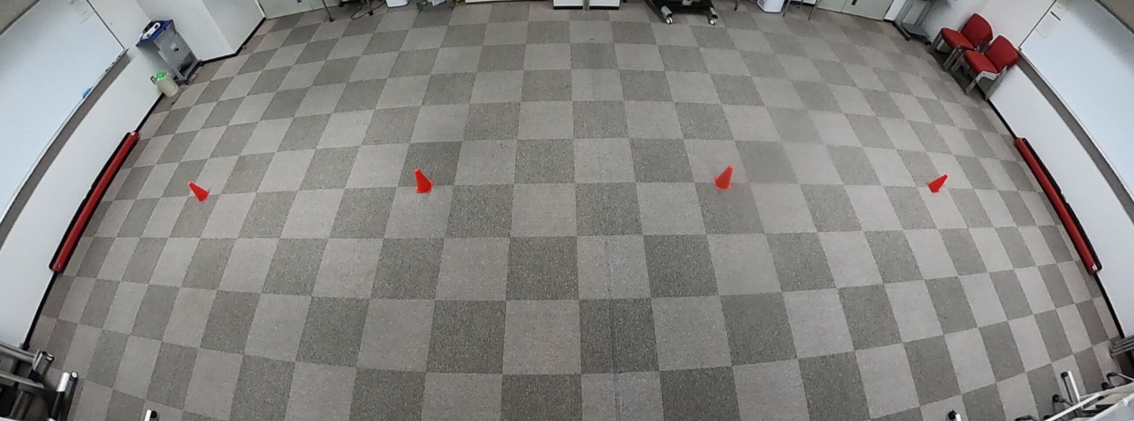}
    \caption{Regular path}
    \label{MD_Re}
  \end{subsubfigure}
  \caption{Manual Driving (MD)}
  \label{fig:MD}
\end{subfigure}

\caption{Experimental environment with all conditions for each path}
\label{fig:condition}
\end{figure}

\subsection{Driving Conditions}

For each of the two driving paths, three driving conditions were designed for the experiment as shown in Fig.~\ref{fig:condition}, \ie manual driving (MD), automated driving without indication of path (AD w/o path) and automated driving with indication of path (AD w/ path).

\subsubsection{Automated driving without path indication (AD w/o path)}

In the AD w/o path condition, no information regarding the driving path was presented in the experimental site.
Participants rode the APMV in empty rooms under two path conditions (see Fig.~\ref{fig:AD w/o path}).
Participants were asked to sit in a relaxed posture on the APMV and experience automated driving.
The APMV was set with a maximum linear velocity of $6~\mathrm{km/h}$ and a maximum linear acceleration of $1.7~\mathrm{m/s}^2$, and its specific movements were controlled in real time by the autonomous driving system.

\subsubsection{Automated driving with path indication (AD w/ path)}

Consistent with the AD w/o path condition, participants experienced automated driving by the autonomous driving system with a maximum linear velocity of $6~\mathrm{km/h}$ and a maximum linear acceleration of $1.7~\mathrm{m/s}^2$.
The difference from the AD w/o path condition was that the driving path was presented to the participants using green tape to ensure clear visibility (see Fig.~\ref{fig:AD w/ path}).
Participants were asked to sit in a relaxed posture on the APMV and experience automated driving.
They were informed that the green tape represented the driving path, but they were not required to fix their gaze on it throughout the ride.

\subsubsection{Manual driving (MD)}

Participants were informed that this condition simulated manual driving of the APMV in a shared space.
Some pylons were placed in the experimental site to simulate pedestrians (see Fig.~\ref{fig:MD}). 
Participants were instructed to manually drive the APMV themselves while avoiding these pylons.
Specifically, in the irregular path condition, participants were instructed to pass to the right of the red pylons, to the left of the blue pylons, and to navigate halfway around the green pylons from the left side (see Fig.~\ref{MD_Ire}).
In the regular path condition, red pylons were placed at fixed intervals, and participants were instructed to drive in a slalom pattern between them (see Fig.~\ref{MD_Re}).
Consistent with the two AD conditions, the same maximum velocity and maximum linear acceleration limits were applied, while the driving behavior was manually controlled by each participant.

\subsection{Participants}

An \textit{a priori} power analysis was performed using \textit{G*Power} (Version~3.1.9.7~\citep{GPower}) for a mixed-design ANOVA with two between-subject factors (genders: 2 levels; driving paths: 2 levels) and one within-subject factor (driving conditions: 3 levels). 
The analysis focused on the critical within--between interaction effect corresponding to the Gender $\times$ Path $\times$ Condition interaction.
For the \textit{G*Power}, genders and driving paths were combined into $2 \times 2 = 4$ independent between-subject groups, and driving conditions was specified as 3 repeated measurements.
Thus, the $2 \times 2 \times 3$ design was entered as $4 \times 3$ in \textit{G*Power} .
With an assumed medium effect size $f = 0.25$, $\alpha = 0.05$, statistical power of $1-\beta = 0.80$, the required total sample size was estimated to be $N = 40$ participants.

Therefore, a total of 40 participants (20 males and 20 females, self-reported), aged 22–30 years (mean age = 23.8 years, SD = 1.7 years), were recruited from 
[Anonymous]
through campus-wide recruitment emails and participated in the experiment.
Participants were randomly assigned to the two path conditions in a balanced manner.
Specifically, 10 males and 10 females were allocated to the irregular path group, and the remaining 10 males and 10 females were allocated to the regular path group.
All participants reported being in good physical health and indicated no history of vestibular disorders or related balance impairments.
In addition, none of the participants had ever used APMVs and AVs.
All participants received a detailed explanation of the procedure and completed a consent form.
No participants withdrew during the experiment.
Participants received compensation of 3,000 yen for their participation.

Prior to conducting the experiment, the short version of Motion Sickness Susceptibility Questionnaire (MSSQ-short)~\citep{golding2006predicting} was used to assess the subjects' susceptibility to motion sickness.
As a result, the MSSQ-short mean percentile score of the experimental participants had 60.2\% (SD = 25.2\%).
This indicates that the participants in this experiment had a relatively higher susceptibility to motion sickness.
To determine whether there were significant differences in MSSQ scores among the four groups (Gender: 2 levels; Path: 2 levels) of participants, a Kruskal--Wallis test was conducted, as the Shapiro--Wilk test indicated that MSSQ scores were not normally distributed ($W = 0.936$, $p < .001$). 
The analysis revealed no significant differences among the groups ($H(3) = 4.04$, $p = .258$). 
This suggests that baseline motion sickness susceptibility did not differ across groups.

\subsection{Procedure}

First, the participants were informed about the purpose of the experiment and the potential risk of motion sickness.
After confirming their understanding and willingness to participate, participants were asked to complete a written consent form.
Subsequently, participants were asked to report their current physical state and any previous experience with AVs or APMVs.
Next, the participants took their seats on the APMV, and the experimenter assisted them in wearing a helmet equipped with an IMU.

Each participant was required to ride the APMV under all three driving conditions. 
The total duration for each condition was 20 minutes, consisting of 15 minutes of driving and 5 minutes of parking.
During the 5 minutes parking period, passengers were instructed to remain seated on the APMV and rest.
Additionally, for each participant, each condition was conducted on a different day, allowing sufficient sleep to alleviate any motion sickness symptoms carried over from the previous condition.
Across all adjacent conditions, the median interval was 76 hours ($Q1=24$ hours; $Q3=191$ hours; $range = [15, 886]$ hours).

To avoid exposing information about the driving path before the AD w/o path condition, all participants first experienced the AD w/o path condition, as driving path information was provided in both the MD and AD w/ path conditions.
Furthermore, to minimize the order effects between the MD and AD w/ path conditions, the sequence of these two conditions was counterbalanced via a Latin square design.

\subsection{Measurements and Data Analysis}

\subsubsection{Driving Motions of APMV}

To compare the APMV motion across the three driving conditions, particularly the differences between the two AD conditions and the MD condition, onboard IMU data were collected and analyzed for each of the three conditions under both driving paths.

To quantify driving motion differences, root mean square (RMS) magnitudes of the three-axis linear acceleration $\mathbf{a}=(a_x, a_y, a_z)$ and three-axis angular velocity $\mathbf{\omega}=(\omega_x, \omega_y, \omega_z)$ were calculated for 15 minutes driving phase in each trial based on the onboard IMU recordings.
For each trial, the RMS magnitude was computed separately for linear acceleration and angular velocity by aggregating the respective three axes.
The RMS magnitude was defined as:
\begin{equation}
RMS(\mathbf{a}) =
\sqrt{
\frac{1}{T}
\sum_{t=1}^{T}
\left(
a_{x}^2(t) + a_{y}^2(t) + a_{z}^2(t)
\right)
}
;\\
RMS(\mathbf{\omega}) =
\sqrt{
\frac{1}{T}
\sum_{t=1}^{T}
\left(
\omega_{x}^2(t) + \omega_{y}^2(t) + \omega_{z}^2(t)
\right)
},
\end{equation}
where, $T$ represents the number of samples in the measured time-series data.
These RMS magnitude of trials were statistically compared across the three driving conditions within each driving path.

\subsubsection{Motion Sickness}

To assess motion sickness progression, the Motion 
Illness Symptom Classification (MISC)~\citep{MISC}, an 11-point scale ranging from 0 (no symptoms) to 10 (vomiting), was used (see Table~\ref{tab:MISC}).
Participants verbally reported their MISC scores at one minute intervals throughout the 20 minutes experiment, which included both the ride and parking phases. 
For safety reasons, if a participant reported MISC scores of 6 or above on two consecutive evaluations, the 15 minutes ride phase was stopped and a 5 minutes parking phase was initiated.

To statistically assess participants' motion sickness progressions, four metrics were computed to quantify motion sickness severity and onset latency in each trial:
\begin{enumerate}[label=\arabic*), itemsep=1pt]
    \item \textbf{MISC mean}: the mean of the MISC scores reported throughout the 20 minutes experiment.
    \item \textbf{MISC max}: the maximum MISC scores reported throughout the 20 minutes experiment.
    \item \textbf{MISC T1}: the initial time required to reach or exceed a MISC score of 1.
    \item \textbf{MISC T2}: the initial time required to reach or exceed a MISC score of 2.
\end{enumerate}
MISC mean and MISC max represent overall and peak motion sickness severity, respectively, whereas MISC T1 and MISC T2 indicate onset latency.
Here, if the participant's MISC scores are all zero during the 20 minutes experiment, then the MISC T1 and MISC T2 are set to 20 minutes.

\begin{figure}[!b]

\centering
\captionof{table}{Motion 
Illness Symptom Clasification(MISC)~\citep{MISC} in English and Japanese was used to assess subjective symptoms of motion sickness in an 11-point scale.}
\label{tab:MISC}

\begin{tabular}{p{7.5cm}|c}
\toprule
\textbf{Symptom} & \textbf{Score} \\
\midrule
No Problem \begin{CJK}{UTF8}{ipxm}不快感なし\end{CJK} & 0 \\
\midrule
\begin{tabular}[c]{@{}l@{}}
Some discomfort but no specific symptons\\
\begin{CJK}{UTF8}{ipxm}違和感・不快感があるが，特定の症状はない\end{CJK} \end{tabular}& 1 \\
\midrule
\begin{tabular}[c]{@{}l@{}}
Have symptoms but no nausea \\ 
\begin{CJK}{UTF8}{ipxm}症状はあるが，吐き気はなし\end{CJK}:\\
{\footnotesize Dizziness, cold/warm, headache, yawning, stomach/throat}\\
{\footnotesize awareness, sweating, blurred vision, eye fatigue, burping,} \\
{\footnotesize tiredness, salivation, etc. }\\ 
{\footnotesize\begin{CJK}{UTF8}{ipxm}めまい，寒気，ほてり，頭痛，あくび，胃やのどの違和感，\end{CJK}}\\
{\footnotesize\begin{CJK}{UTF8}{ipxm}発汗，眼精疲労，げっぷ，疲労，唾液の分泌など\end{CJK} }\end{tabular} &
\begin{tabular}[c]{@{}l@{}}2: Vague (\begin{CJK}{UTF8}{ipxm}微弱\end{CJK}) \\
3: Slight (\begin{CJK}{UTF8}{ipxm}弱\end{CJK}) \\ 
4: Rather (\begin{CJK}{UTF8}{ipxm}中\end{CJK}) \\ 
5: Severe (\begin{CJK}{UTF8}{ipxm}強\end{CJK})\end{tabular} 
\\
\midrule
Nausea \begin{CJK}{UTF8}{ipxm}むかつき・吐き気はあり\end{CJK} &
\begin{tabular}[c]{@{}l@{}}6: Slight (\begin{CJK}{UTF8}{ipxm}弱\end{CJK}) \\ 7: Fairly (\begin{CJK}{UTF8}{ipxm}中\end{CJK}) \\ 8: Severe (\begin{CJK}{UTF8}{ipxm}強\end{CJK}) \\ 9: Retching nausea\\  \ \ \  (\begin{CJK}{UTF8}{ipxm}吐きそう\end{CJK})\end{tabular} \\
\midrule
Vomiting \begin{CJK}{UTF8}{ipxm}嘔吐\end{CJK}& 10 \\
\bottomrule
\end{tabular}
\end{figure}

A nonparametric linear mixed-effects model based on the Aligned Rank Transform (ART)~\citep{wobbrock2011aligned} procedure was used (ARTool package in R) to analyze the MISC mean, MISC max, MISC T1 and MISC T2, separately.
The model included the driving conditions (Condition), driving paths (Path) and passenger genders (Gender) as fixed effects and participant (ID) as a random intercept:
\begin{equation}
\text{DV} \sim \text{Condition} \times \text{Path} \times \text{Gender} + (1 \mid \text{ID}),\ \
\text{where } \text{DV} \in \{ \text{MISC mean}, \text{MISC max}, \text{MISC T1},\text{MISC T2} \}. \nonumber
\end{equation}
When significant main or interaction effects were detected, post-hoc pairwise comparisons were performed using ART-C~\citep{elkin2021aligned}, with p-values adjusted via false discovery rate (FDR) method.

\subsubsection{Delay time between the passengers' head motion and APMV motion}

During the 20 minutes APMV ride, participants' head motion were recorded using an IMU attached to the helmet, and the APMV's motion data were captured using an onboard IMU. 
Both IMUs measured 3-dof of linear acceleration and 3-dof of angular velocity at a sampling rate of 100 Hz.
All data were collected via ROS and stored in \textit{rosbag} files. 
Although both IMUs shared a common system time base, their sampling timestamps were not perfectly aligned. 
Therefore, to correct for asynchronous sampling and achieve temporal alignment between the two IMU signals, zero-order interpolation was applied, and the data were resampled at 100 Hz, following~\citep{liu2024subjective}.

\begin{figure}[b]
  \centering
  \begin{subfigure}{\linewidth}
    \centering
    \includegraphics[width=0.8\linewidth]{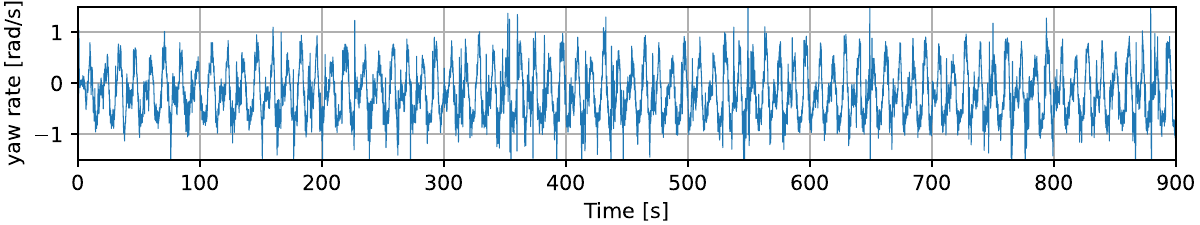}
     \vspace{-1mm}
    \caption{Yaw rate of head}
  \end{subfigure}\\
  \begin{subfigure}{\linewidth}
    \centering
    \includegraphics[width=0.8\linewidth]{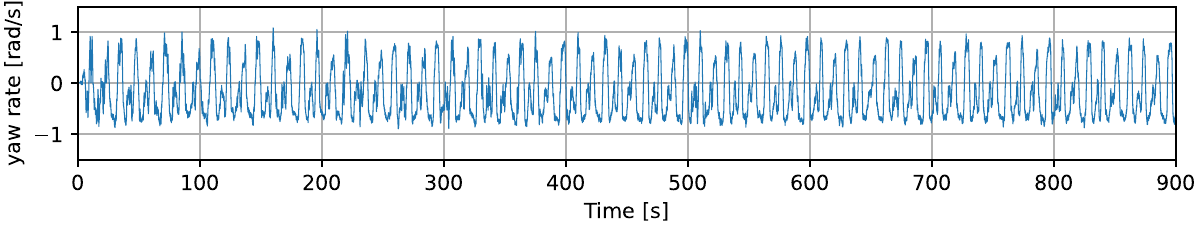}
    \vspace{-1mm}
    \caption{Yaw rate of APMV}
  \end{subfigure}\\
  \begin{subfigure}{\linewidth}
    \centering
    \includegraphics[width=0.9\linewidth]{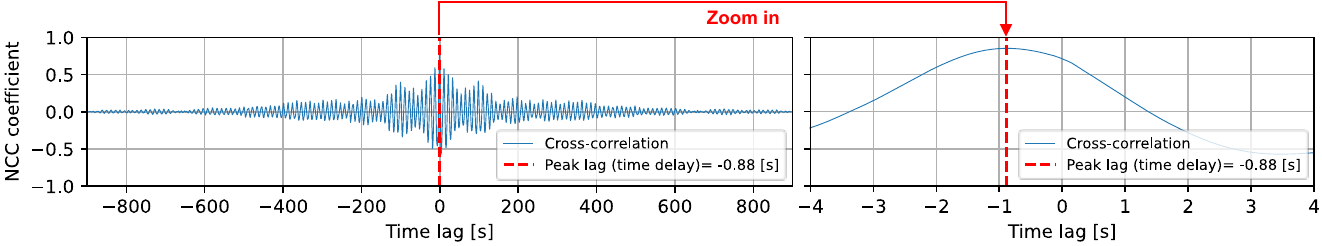}
    \vspace{-1mm}
    \caption{The peak lag is estimated as the delay time via normalized cross-correlation between yaw rates of head and APMV.}
  \end{subfigure}
  \vspace{-6mm}
  \caption{Example of delay time estimation via normalized cross-correlation for participant i12 under MD on irregular path.}
  \label{fig:vertical-captions}
\end{figure}

To examine whether indicating the driving path was associated with compensatory behavior in passengers, referred to \citep{wijlens2025road}, normalized cross-correlation (NCC) was used to estimate the time lag of the participant's head motion relative to the corresponding APMV motion during 15 minutes ride, using the APMV motion as the reference 
(\ie, the delay of head motion in the same direction as the APMV).
Since the APMV used in this experiment was not equipped with a suspension system, pitch and roll rotations were negligible during driving on a flat surface.
In addition, compared with conventional cars, the APMV generated relatively small lateral accelerations during turning. 
Therefore, the NCC analysis was conducted using yaw rate, which represented the dominant motion component of the APMV and passengers' head.
The function of NCC is:  
\begin{equation}
NCC(\tau) = \frac{\sum_{t=\max(1, 1-\tau)}^{\min(T, T-\tau)} \left(\omega_x^{(APMV)}(t) - \bar{\omega}_x^{(APMV)} \right) \left(\omega_x^{(Head)}(t + \tau) - \bar{\omega}_x^{(Head)} \right)}{\sqrt{\sum_{t=\max(1, 1-\tau)}^{\min(T, T-\tau)} \left(\omega_x^{(APMV)}(t) - \bar{\omega}_x^{(APMV)} \right)^2 \sum_{t=\max(1, 1-\tau)}^{\min(T, T-\tau)} \left(\omega_x^{(Head)}(t + \tau) - \bar{\omega}_x^{(Head)} \right)^2}}
\end{equation}
where, $\omega_x^{(APMV)}(t)$ and $\omega_x^{(Head)}(t)$ denote the yaw rate of the APMV and participant's head.
Their mean yaw rates over time are denoted by $\bar{\omega}_x^{(APMV)}$ and $\bar{\omega}_x^{(Head)}$, respectively. 
$T$ represents the number of samples in the measured time-series data, and $\tau\in [-T, T]$ denotes the time lag.  
The time lag $\tau$ corresponding to the maximum positive coefficient of NCC was regarded as the estimated delay time.
A negative delay time indicates that the participant's head rotation preceded that of the APMV, whereas a positive value indicates a lag in head rotation relative to the APMV.
Fig.~\ref{fig:vertical-captions} presents an example of the two IMUs data collected during a 15 minutes ride in which participant \#12 manually operated the APMV (MD condition), along with the estimated delay time obtained by NCC.
In this example, the estimated delay time was –0.88 (s), indicating that the participant \#12's head rotated 0.88 second earlier than the APMV.

As same as the MISC metrics, a nonparametric linear mixed-effects model based on the ART was used to analyze the delay time:
\begin{equation}
\text{Delay time} \sim \text{Condition} \times \text{Path} \times \text{Gender} + (1 \mid \text{ID}).\nonumber
\end{equation}
Post-hoc pairwise comparisons were performed using ART-C with FDR adjusted p-values, when significant main or interaction effects were detected.

\section{RESULTS}

All 40 participants completed the full experiment. No participant withdrew during the experiment, and no MISC score exceeded 6 for any participant.
The following results are based on data from all 40 participants.

\subsection{Driving Motions of APMV}

\begin{figure}[b]
\centering
  \begin{subfigure}{0.32\textwidth}
    \centering
    \includegraphics[width=1\linewidth]{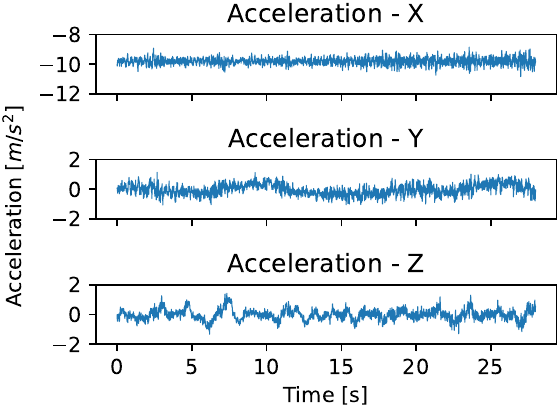}
    \caption{Acceleration of APMV under the MD condition}
\end{subfigure}
\hfill
\begin{subfigure}{0.32\textwidth}
    \centering
    \includegraphics[width=1\linewidth]{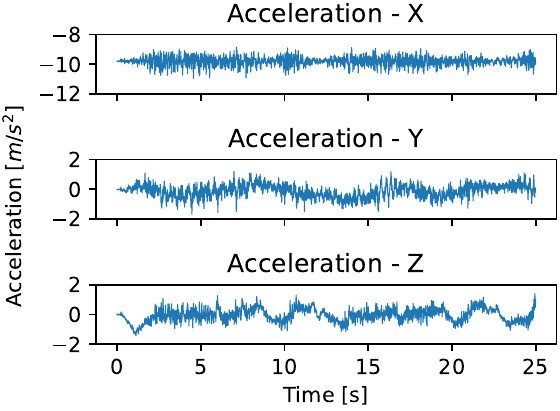}
    \caption{Acceleration of APMV under the AD w/o path condition}
\end{subfigure}
\hfill
\begin{subfigure}{0.32\textwidth}
    \centering
    \includegraphics[width=1\linewidth]{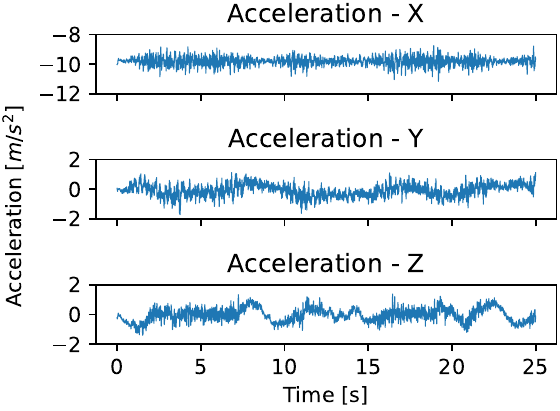}
    \caption{Acceleration of APMV under the AD w/ path condition}
\end{subfigure}
\\
 \begin{subfigure}{0.32\textwidth}
    \centering
    \includegraphics[width=1\linewidth]{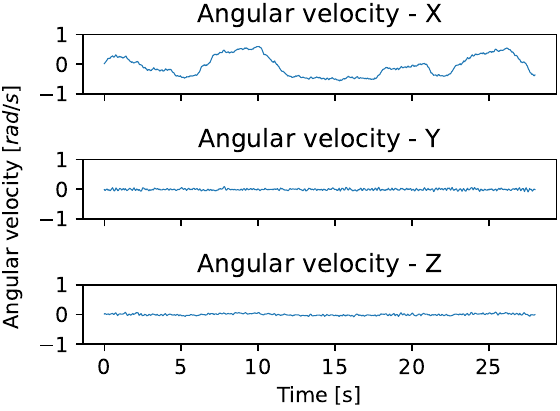}
    \caption{Angular velocity of APMV under the MD condition}
\end{subfigure}
\hfill
\begin{subfigure}{0.32\textwidth}
    \centering
    \includegraphics[width=1\linewidth]{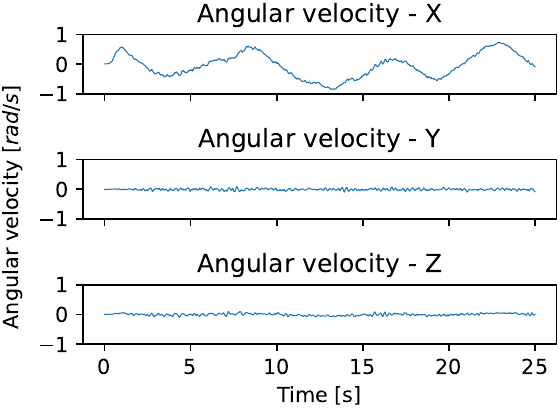}
    \caption{Angular velocity of APMV under the AD w/o path condition}
\end{subfigure}
\hfill
\begin{subfigure}{0.32\textwidth}
    \centering
    \includegraphics[width=1\linewidth]{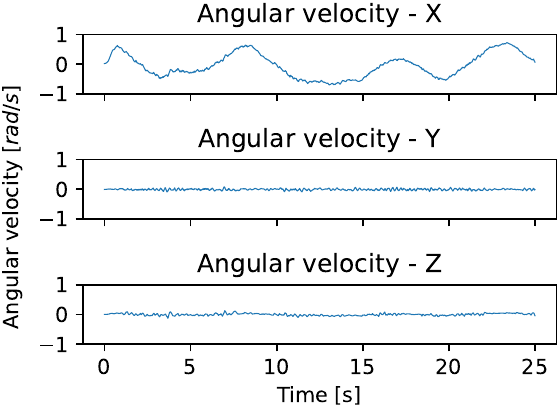}
    \caption{Angular velocity of APMV under the AD w/ path condition}
\end{subfigure}
\caption{Representative example of APMV motion recorded by the onboard IMU during a single loop on the irregular path (Participant i03).}
    \label{fig:IMU_i03}
    \vspace{-1mm}
\end{figure}

\begin{figure}[!h]
\centering
  \begin{subfigure}{0.32\textwidth}
    \centering
    \includegraphics[width=1\linewidth]{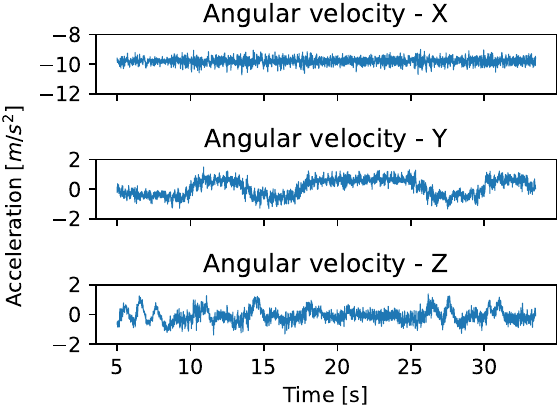}
    \caption{Acceleration of APMV under the MD condition}
\end{subfigure}
\hfill
\begin{subfigure}{0.32\textwidth}
    \centering
    \includegraphics[width=1\linewidth]{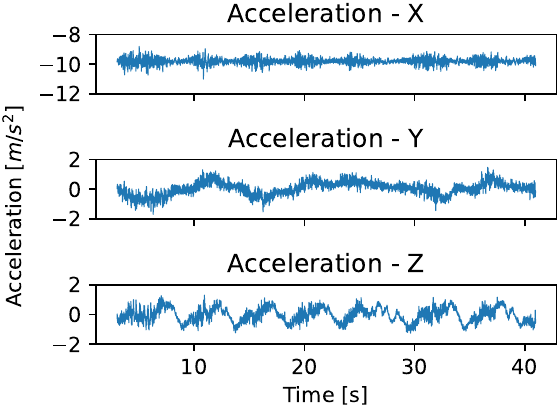}
    \caption{Acceleration of APMV under the AD w/o path condition}
\end{subfigure}
\hfill
\begin{subfigure}{0.32\textwidth}
    \centering
    \includegraphics[width=1\linewidth]{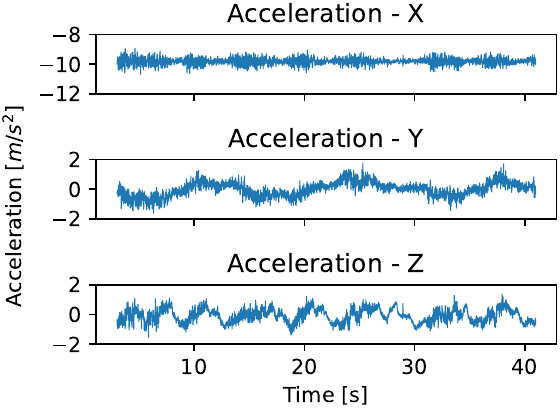}
    \caption{Acceleration of APMV under the AD w/ path condition}
\end{subfigure}
\\
 \begin{subfigure}{0.32\textwidth}
    \centering
    \includegraphics[width=1\linewidth]{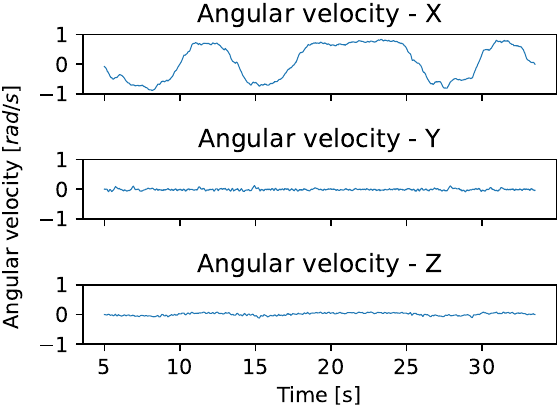}
    \caption{Angular velocity of APMV under the MD condition}
\end{subfigure}
\hfill
\begin{subfigure}{0.32\textwidth}
    \centering
    \includegraphics[width=1\linewidth]{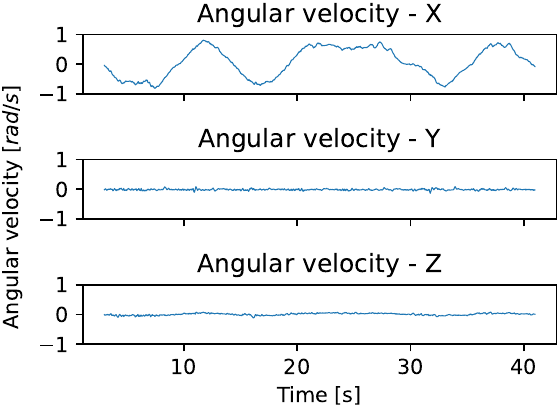}
    \caption{Angular velocity of APMV under the AD w/o path condition}
\end{subfigure}
\hfill
\begin{subfigure}{0.32\textwidth}
    \centering
    \includegraphics[width=1\linewidth]{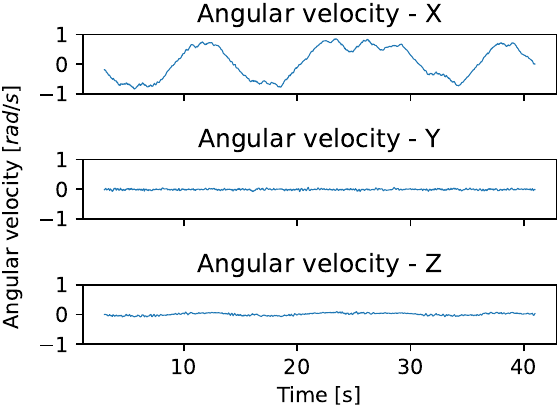}
    \caption{Angular velocity of APMV under the AD w/ path condition}
\end{subfigure}
\caption{Representative example of APMV motion recorded by the onboard IMU during a single loop on the regular path (Participant r05).}
    \label{fig:IMU_r05}
\vspace{-5mm}
\end{figure}

Representative examples of APMV motion recorded by the onboard IMU are shown in Figs.~\ref{fig:IMU_i03} and~\ref{fig:IMU_r05} for the irregular and regular paths, respectively.
These figures show three-axis linear acceleration and three-axis angular velocity under the three driving conditions (MD, AD w/o path, and AD w/ path) during a single loop.

\begin{figure}[!t]
  \centering
\begin{subfigure}{0.24\textwidth}
    \centering
    \includegraphics[width=1\linewidth]{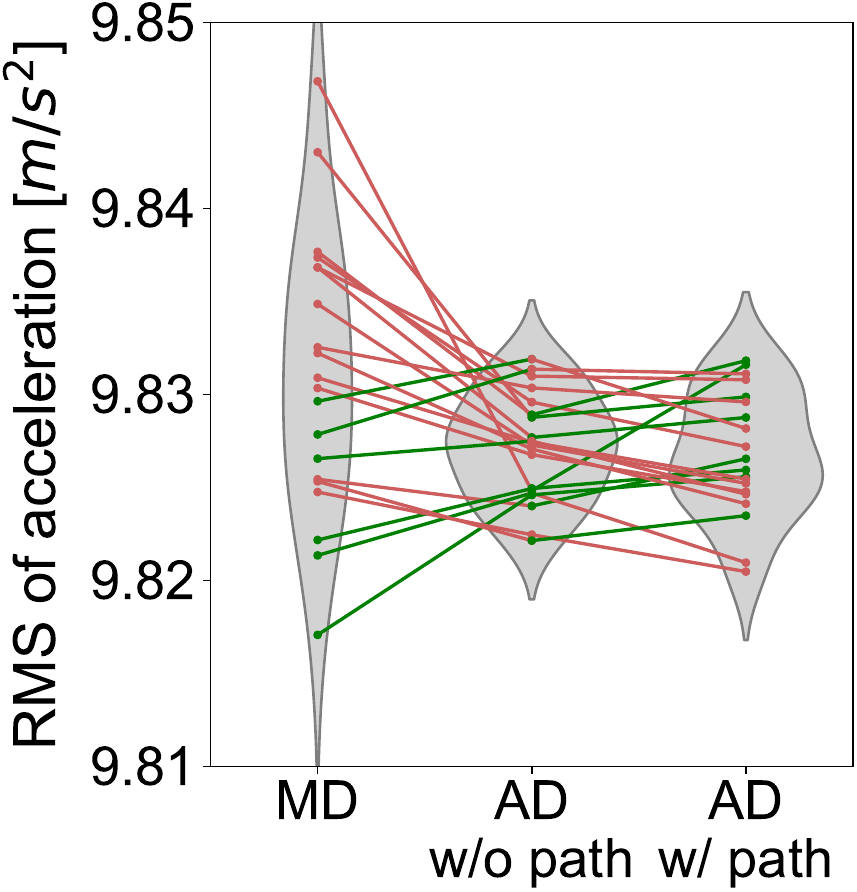}
    \caption{RMS of accelerations on the irregular path}
    \label{fig:RMS_ACC_Irr}
\end{subfigure}
\hfill
\begin{subfigure}{0.24\textwidth}
    \centering
    \includegraphics[width=1\linewidth]{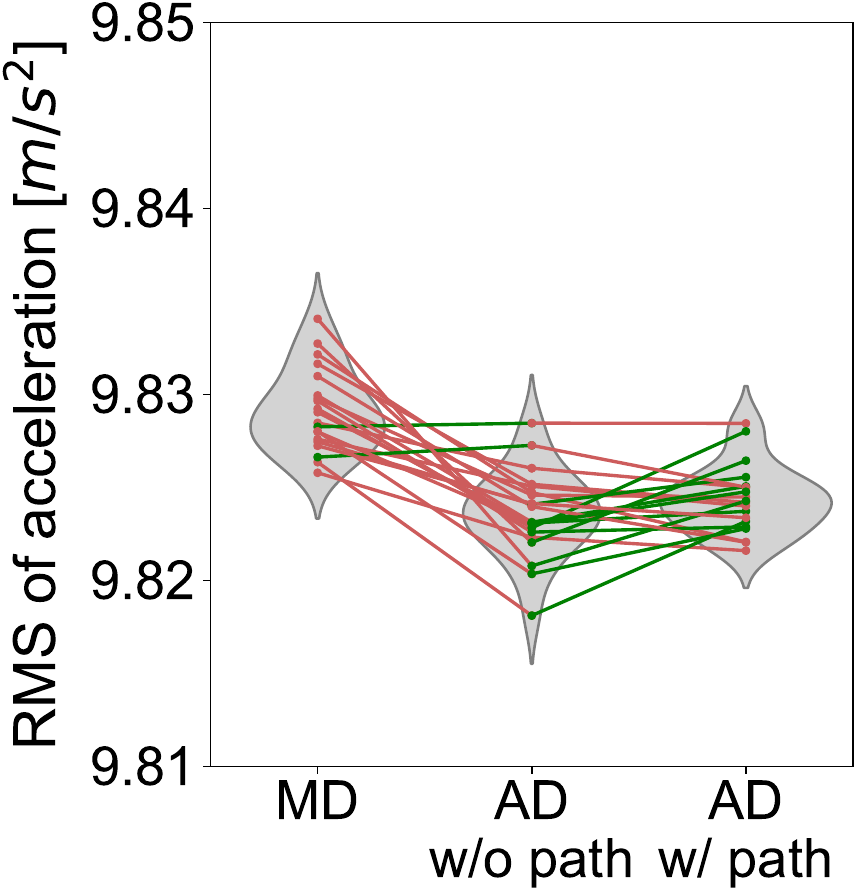}
    \caption{RMS of accelerations on the regular path}
    \label{fig:RMS_ACC_Re}
\end{subfigure}
\hfill
\begin{subfigure}{0.24\textwidth}
    \centering
    \includegraphics[width=1\linewidth]{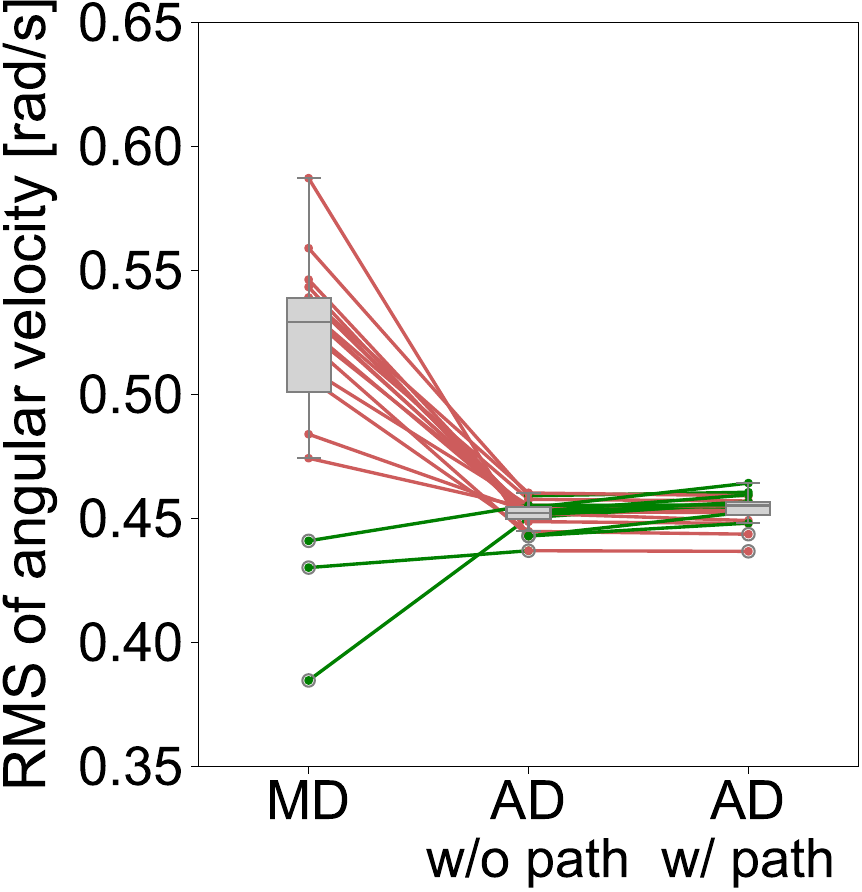}
    \caption{RMS of angular velocity on the irregular path}
     \label{fig:RMS_Ang_Irr}
\end{subfigure}
\hfill
\begin{subfigure}{0.24\textwidth}
    \centering
    \includegraphics[width=1\linewidth]{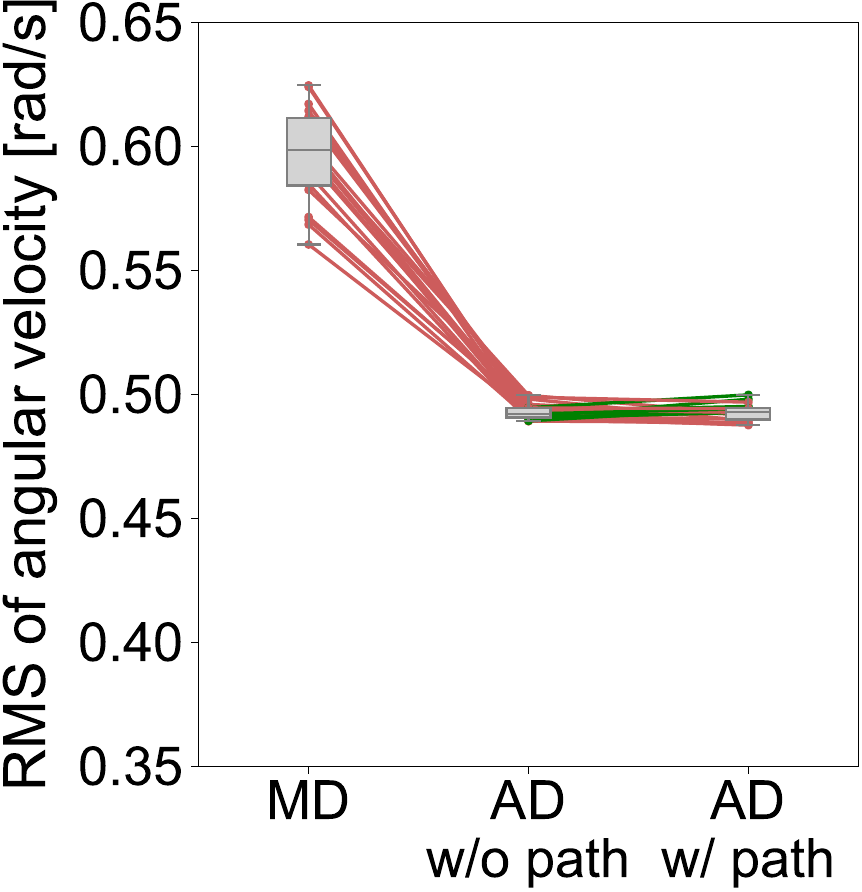}
    \caption{RMS of angular velocity on the regular path}
    \label{fig:RMS_Ang_Re}
\end{subfigure}
 \caption{Paired violin and box plots of RMS of acceleration and angular velocity across three conditions on the two driving paths.
 The violin plots and box plots show the distribution of RMS.
Color lines represent within-subject changes between adjacent conditions. Green lines indicate an increase from left to right, red lines a decrease, and gray lines no change.}
  \label{fig:RMS}

\end{figure}

To quantitatively evaluate differences in APMV motion among the three driving conditions, the RMS of of linear acceleration and angular velocity was calculated.
For RMS of accelerations on both the irregular and regular paths (see Figs.~\ref{fig:RMS_ACC_Irr} and ~\ref{fig:RMS_ACC_Re}), normality was confirmed using the Shapiro–Wilk test (all $p > .05$), and Mauchly's test indicated that the sphericity assumption was satisfied.
Separate one-way repeated-measures ANOVAs revealed a significant main effect of condition on RMS of accelerations for both paths (irregular: \textit{F}(2, 38) = 7.36, \textit{p} =.002, $\eta^2_g$ = .13; regular: \textit{F}(2, 38) = 43.92, \textit{p} < .001, $\eta^2_g$ = .58.).
Post-hoc paired \textit{t}-tests with Holm-Bonferroni correction showed that, for both driving paths, the RMS of accelerations under the MD condition was significantly higher than it under the both AD conditions respectively (all adjusted \textit{p} < .001), whereas no significant difference was observed between the two AD conditions.

For RMS of angular velocity on both the irregular and regular paths (see Figs.~\ref{fig:RMS_Ang_Irr} and ~\ref{fig:RMS_Ang_Re}), the normality assumption was not satisfied in all conditions, a Friedman test was conducted for both paths.
Separate Friedman test revealed a significant main effect of condition on RMS of angular velocity for both paths (irregular: $\chi^2(2) = 14.8$, \textit{p} < .001, $W$ = .37; regular: $\chi^2(2) = 30.4$, \textit{p} < .001, $W$ = .76).
Post-hoc pairwise comparisons using Wilcoxon signed-rank tests with Benjamini-Hochberg FDR correction showed that, for both driving paths, the RMS of angular velocity under the MD condition was significantly higher than it under the both AD conditions respectively (all adjusted \textit{p} < .001), whereas no significant difference was observed between the two AD conditions.

\subsection{MISC metrics}  

\begin{figure}[b]
\centering
  \begin{subfigure}{0.495\textwidth}
    \centering
    \includegraphics[width=1\linewidth]{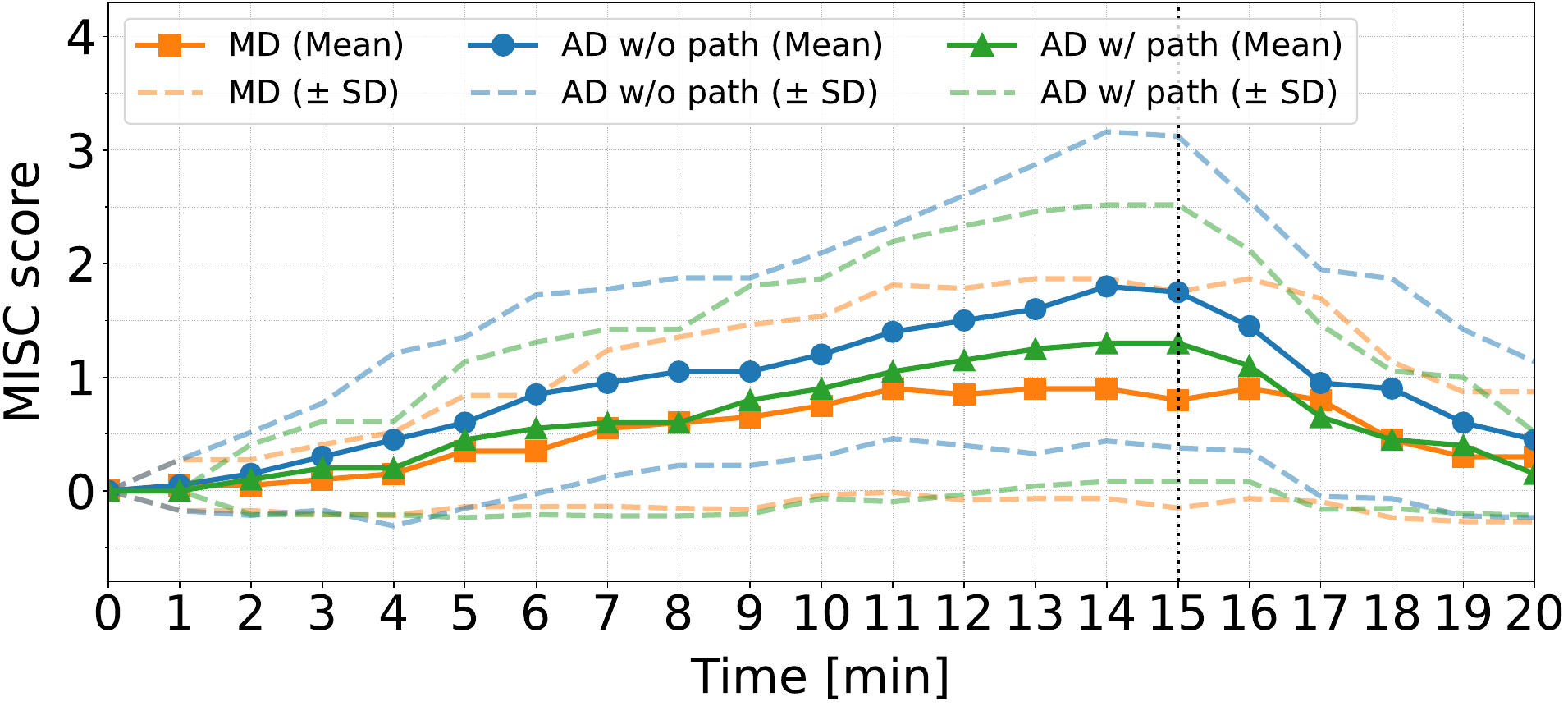}
    \caption{MISC progression of irregular path group ($N=20$)}
\end{subfigure}
\begin{subfigure}{0.495\textwidth}
    \centering
    \includegraphics[width=1\linewidth]{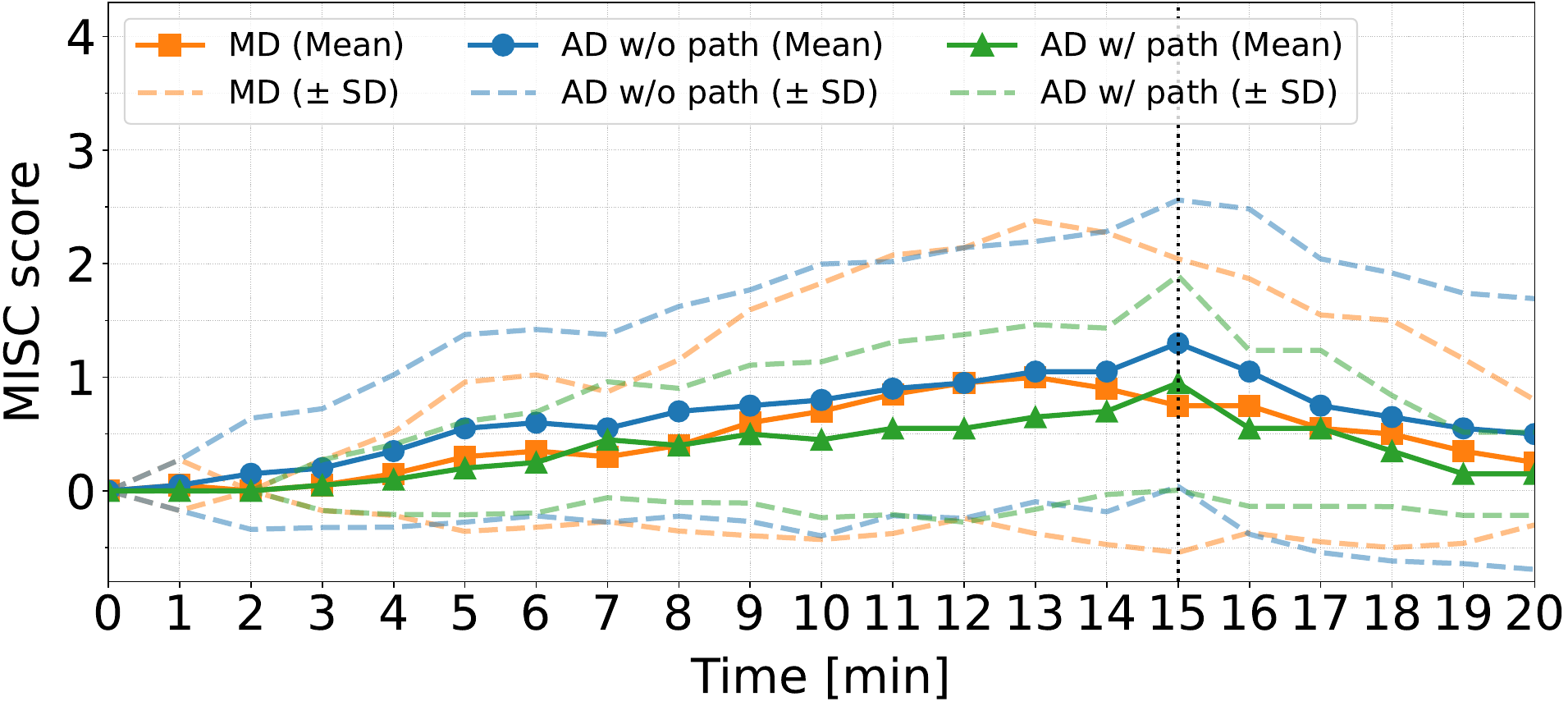}
    \caption{MISC progression of regular path group ($N=20$)}
\end{subfigure}
\caption{MISC progression under each condition.
Solid lines show the mean MISC score at each time point across all participants, and dashed lines show $\pm 1$ SD. The black vertical dotted line at 15 minutes shows the APMV stopping time.}
    \label{fig:MISC_progression}
   \end{figure}

Fig.~\ref{fig:MISC_progression} shows the mean MISC scores and $\pm 1$ SD over the 20 minutes under each condition for both the irregular and regular path groups. 
Overall, MISC scores increased gradually during the ride, reaching a peak shortly before the stopping point at 15 minutes, followed by a decrease after the APMV stopped.

 \begin{figure}[b]
  \centering

\begin{subfigure}{0.24\textwidth}
    \centering
    \includegraphics[width=1\linewidth]{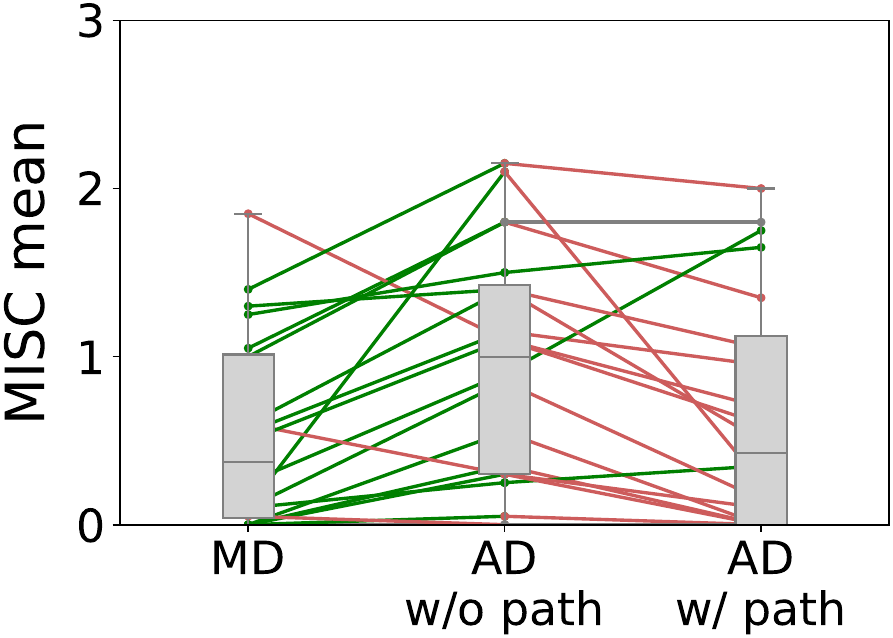}
    \caption{MISC mean of irregular path group}
\end{subfigure}
\hfill
\begin{subfigure}{0.24\textwidth}
    \centering
    \includegraphics[width=1\linewidth]{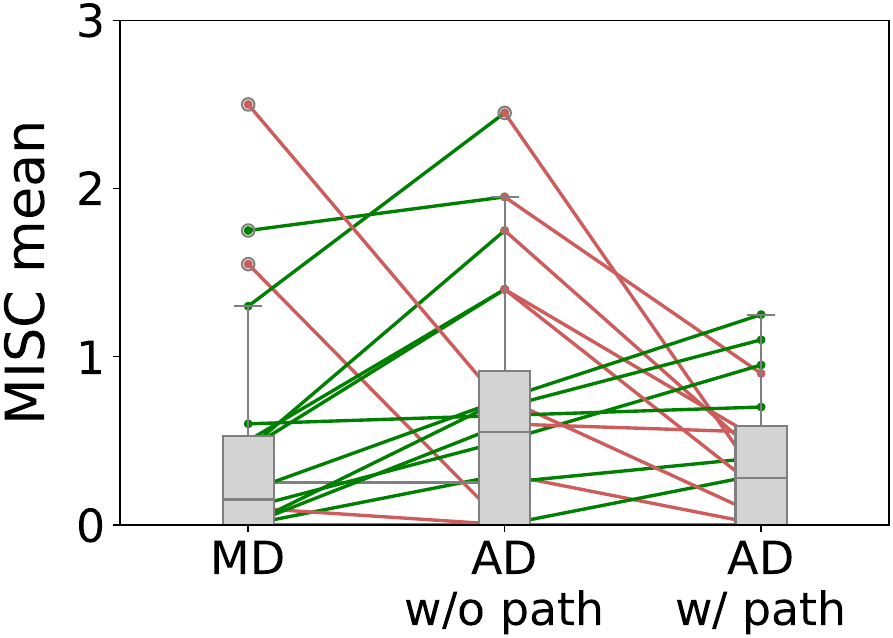}
    \caption{MISC mean of regular path group}
\end{subfigure}
\hfill
\begin{subfigure}{0.24\textwidth}
    \centering
    \includegraphics[width=0.98\linewidth]{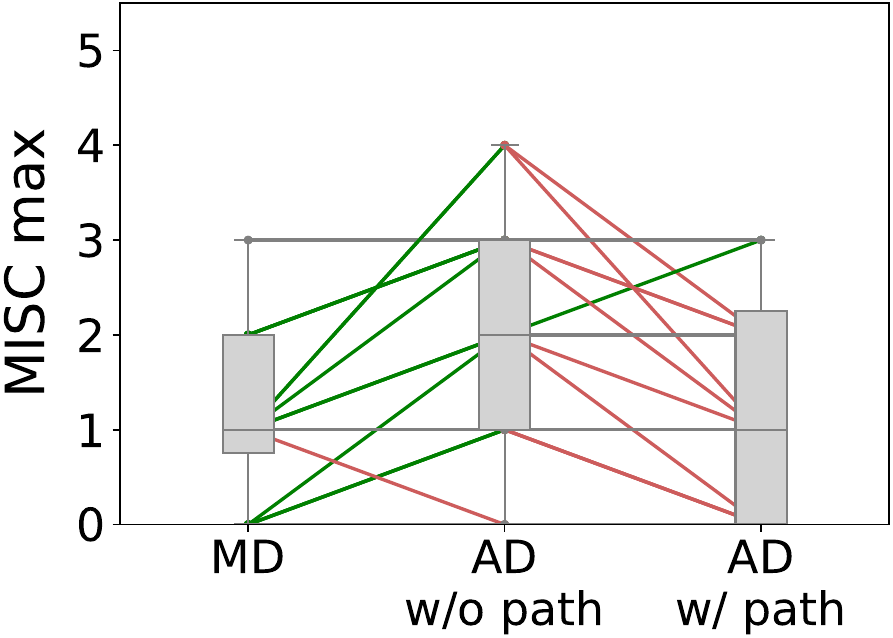}
    \caption{MISC max of irregular path group}
\end{subfigure}
\hfill
\begin{subfigure}{0.24\textwidth}
    \centering
    \includegraphics[width=0.98\linewidth]{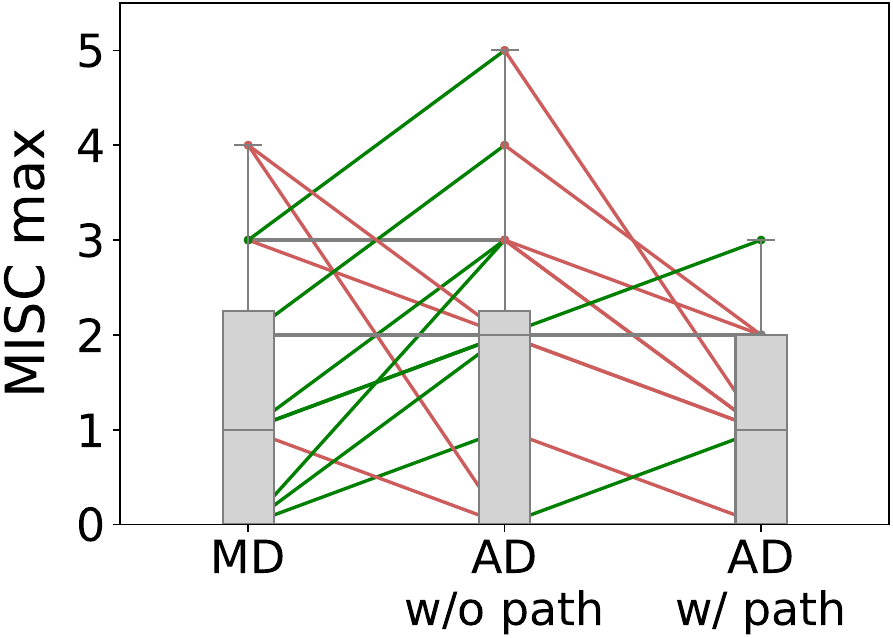}
    \caption{MISC max of regular path group}
\end{subfigure}

\begin{subfigure}{0.24\textwidth}
    \centering
    \includegraphics[width=1\linewidth]{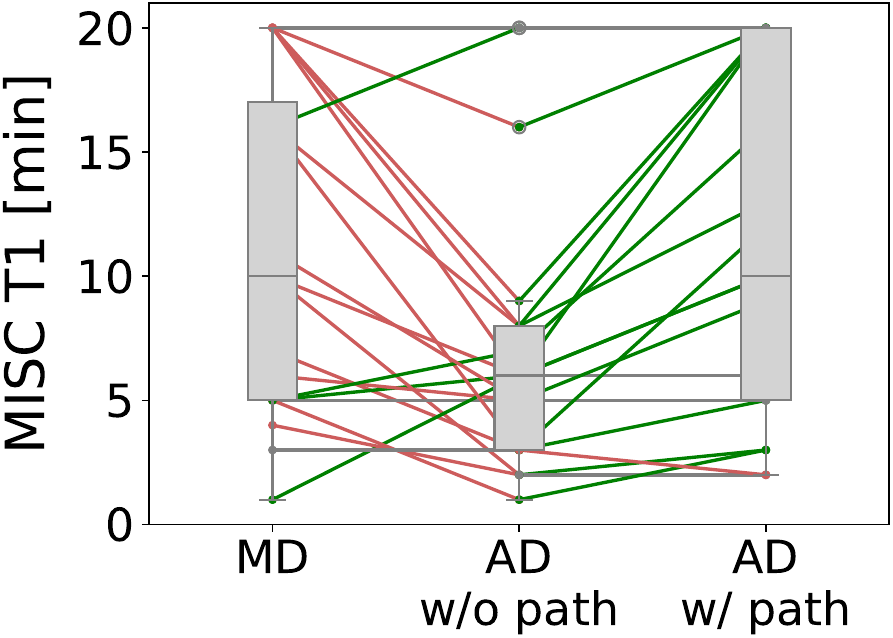}
    \caption{MISC T1 of irregular path group}
\end{subfigure}
\hfill
\begin{subfigure}{0.24\textwidth}
    \centering
    \includegraphics[width=1\linewidth]{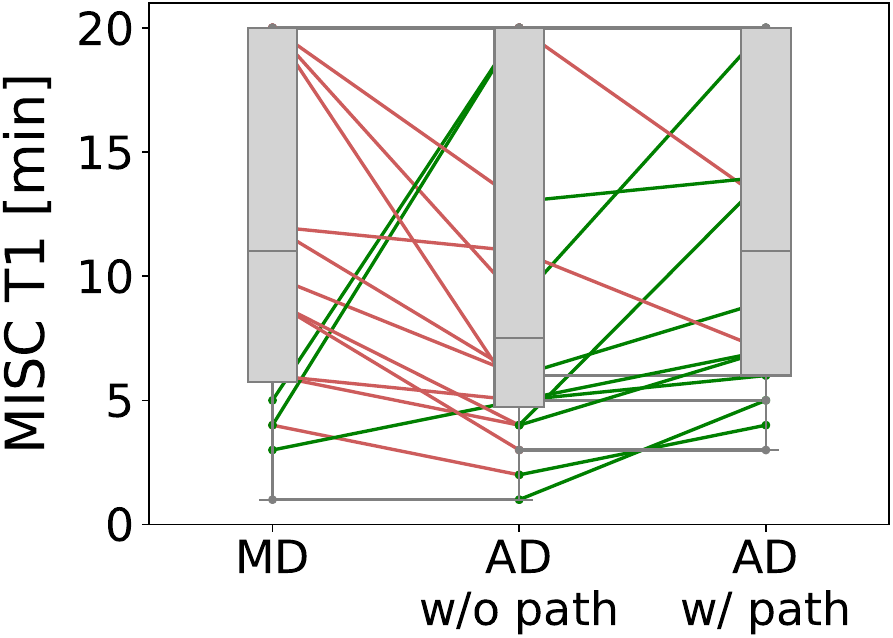}
    \caption{MISC T1 of regular path group}
\end{subfigure}
\hfill
\begin{subfigure}{0.24\textwidth}
    \centering
    \includegraphics[width=1\linewidth]{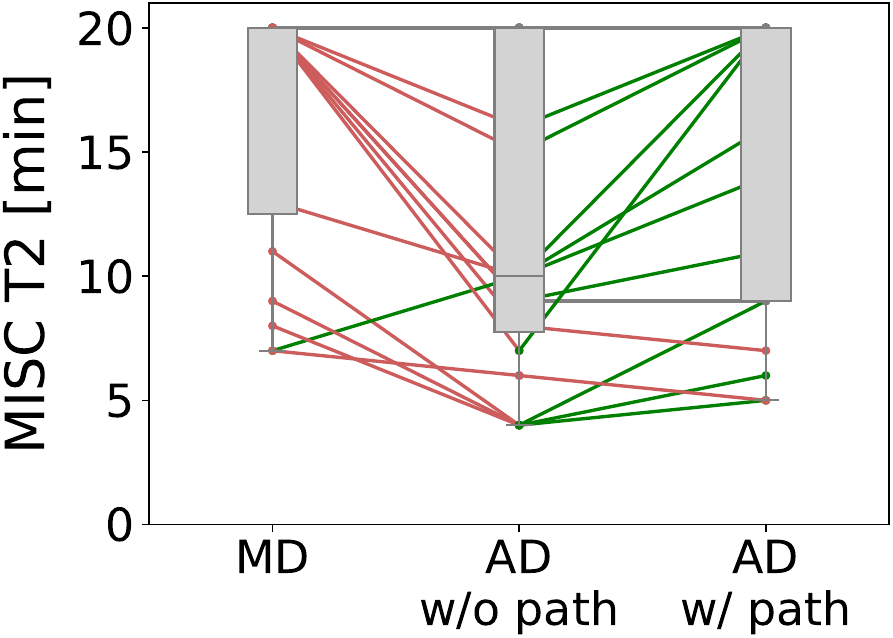}
    \caption{MISC T2 of irregular path group}
\end{subfigure}
\hfill
\begin{subfigure}{0.24\textwidth}
    \centering
    \includegraphics[width=1\linewidth]{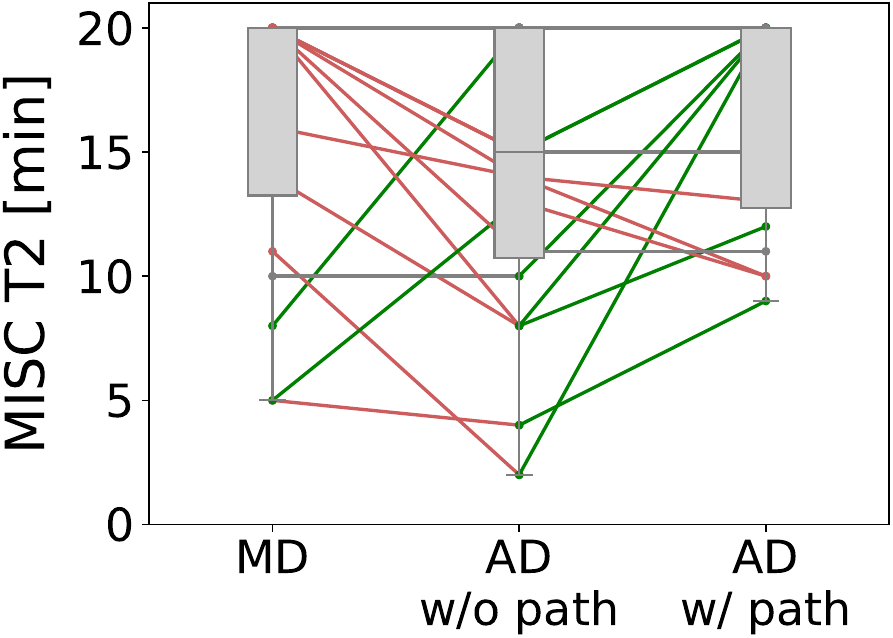}
    \caption{The MISC T2 of regular path group}
\end{subfigure}

  \caption{Paired box plots of MISC metrics.
 The box plots show the distribution of MISC metrics for each condition.
Color lines represent within-subject changes between adjacent conditions. Green lines indicate an increase from left to right, red lines a decrease, and gray lines no change.}
  \label{fig:MISC_results}
\end{figure}

To quantitatively assess differences in MISC among the driving conditions, four MISC metrics, \ie MISC max, MISC mean, MISC T1, and MISC T2, were computed as shown in Fig.~\ref{fig:MISC_results}.
The three-way ART ANOVA revealed a significant main effect of driving conditions for all MISC metrics as shown in Table~\ref{tab:ANOVA-MISC}. 
No significant main effects of Path or Gender were found for any MISC index. 
Additionally, none of the two-way or three-way interaction effects reached statistical significance.
Post hoc ART-C comparisons with FDR adjustment further clarified the significant main effect of driving conditions, as shown in Table~\ref{tab:posthoc-MISC}. 
For all MISC metrics, AD w/o path resulted in significantly higher motion sickness compared to both AD w/ path and MD. 
In contrast, no significant differences were observed between AD w/ path and MD.

\begin{table}[!b]
\footnotesize
\centering
\caption{Results of the three-way ART ANOVA for MISC metrics based on linear mixed-effects models}
\label{tab:ANOVA-MISC}
\begin{tabular}{lrccrr@{\hspace{2pt}}l}
\toprule
Dependent Variable & Source & dof & dof.res & \multicolumn{1}{c}{$F$} & \multicolumn{2}{c}{$p$} \\ \midrule
 \multirow{7}{*}{MISC mean} 
 & Gender & 1 & 36 & 0.087 & .770 &  \\
 & Path & 1 & 36 & 1.244 & .272 &  \\
 & Condition & 2 & 72 & 7.504 & $.001$ & $**$ \\
 & Gender $\times$ Path & 1 & 36 & 0.057 & .813 &  \\
 & Gender $\times$ Condition & 2 & 72 & 0.738 & .482 &  \\
 & Path $\times$ Condition & 2 & 72 & 1.074 & .347 &  \\
 & Gender $\times$ Path $\times$ Condition & 2 & 72 & 0.363 & .697 &  \\ \midrule
 \multirow{7}{*}{MISC max} 
 & Gender & 1 & 36 & 0.098 & .755 &  \\
 & Path & 1 & 36 & 1.167 & .287 &  \\
 & Condition & 2 & 72 & 6.267 & .003 & $**$ \\
 & Gender $\times$ Path & 1 & 36 & 0.235 & .631 &  \\
 & Gender $\times$ Condition & 2 & 72 & 0.177 & .838 &  \\
 & Path $\times$ Condition & 2 & 72 & 2.492 & .090 &  \\
 & Gender $\times$ Path $\times$ Condition & 2 & 72 & 0.791 & .457 &  \\ \midrule
 \multirow{7}{*}{MISC T1} 
 & Gender & 1 & 36 & 0.179 & .675 &  \\
 & Path & 1 & 36 & 0.653 & .424 &  \\
 & Condition & 2 & 72 & 7.680 & $< .001$ & $***$ \\
 & Gender $\times$ Path & 1 & 36 & 0.066 & .798 &  \\
 & Gender $\times$ Condition & 2 & 72 & 0.020 & .981 &  \\
 & Path $\times$ Condition & 2 & 72 & 1.132 & .328 &  \\
 & Gender $\times$ Path $\times$ Condition & 2 & 72 & 0.359 & .700 &  \\ \midrule
 \multirow{7}{*}{MISC T2} 
 & Gender & 1 & 36 & 0.096 & .759 &  \\
 & Path & 1 & 36 & 0.798 & .377 &  \\
 & Condition & 2 & 72 & 7.863 & $< .001$ & $***$ \\
 & Gender $\times$ Path & 1 & 36 & 0.167 & .685 &  \\
 & Gender $\times$ Condition & 2 & 72 & 1.002 & .372 &  \\
 & Path $\times$ Condition & 2 & 72 & 2.072 & .133 &  \\
 & Gender $\times$ Path $\times$ Condition & 2 & 72 & 0.195 & .824 &  \\ 
 \bottomrule
\multicolumn{7}{l}{\footnotesize Gender: male and female. Path (driving paths): irregular path and regular path.}\\
\multicolumn{7}{l}{\footnotesize Condition (driving conditions): MD, AD w/o path and AD W/ path.}\\
\multicolumn{7}{l}{\footnotesize $*$: $p<0.05$; $**$: $p<0.01$; $***$: $p<0.001$.}  
\end{tabular}

\centering
\caption{ART-C Post-hoc comparisons with FDR adjustment for the main effect of driving conditions in each MISC index}
\label{tab:posthoc-MISC}
\begin{tabular}{cccrrrrr@{\hspace{2pt}}l}
\toprule
Dependent Variable & Condition A & Condition B & estimate & SE & dof & t.ratio & \multicolumn{2}{c}{$p$-adj}  \\
\midrule

\multirow{3}{*}{MISC mean}
 & AD w/ path   & AD w/o path & -14.8  & 4.64 & 72 & -3.182 & .003  & $**$ \\
 & AD w/ path   & MD          &   1.5  & 4.64 & 72 &  0.323 & .748  &  \\
 & AD w/o path  & MD          &  16.3  & 4.64 & 72 &  3.505 & .002  & $**$ \\
\midrule

\multirow{3}{*}{MISC max}
 & AD w/ path   & AD w/o path & -15.59 & 5.21 & 72 & -2.989 & .006  & $**$ \\
 & AD w/ path   & MD          &   0.78 & 5.21 & 72 &  0.149 & .882  &  \\
 & AD w/o path  & MD          &  16.36 & 5.21 & 72 &  3.138 & .006  & $**$ \\
\midrule

\multirow{3}{*}{MISC T1}
 & AD w/ path   & AD w/o path &  16.01 & 4.65 & 72 &  3.441 & .002  & $**$ \\
 & AD w/ path   & MD          &   0.45 & 4.65 & 72 &  0.097 & .923  &  \\
 & AD w/o path  & MD          & -15.56 & 4.65 & 72 & -3.345 & .002  & $**$ \\
\midrule

\multirow{3}{*}{MISC T2}
 & AD w/ path   & AD w/o path &  14.15 & 5.09 & 72 &  2.782 & .010  & $*$ \\
 & AD w/ path   & MD          &  -5.38 & 5.09 & 72 & -1.057 & .294  &  \\
 & AD w/o path  & MD          & -19.52 & 5.09 & 72 & -3.839 & $< .001$ & $***$ \\
\bottomrule
\multicolumn{9}{l}{\footnotesize SE: standard error of the estimate. $*$: $p<0.05$; $**$: $p<0.01$; $***$: $p<0.001$.}
\end{tabular}
\end{table}

\clearpage

\subsection{Delay time between the passengers' head motion and APMV motion}

\begin{figure}[h]

\centering
\begin{subfigure}{0.45\textwidth}
    \centering
    \includegraphics[width=0.9\linewidth]{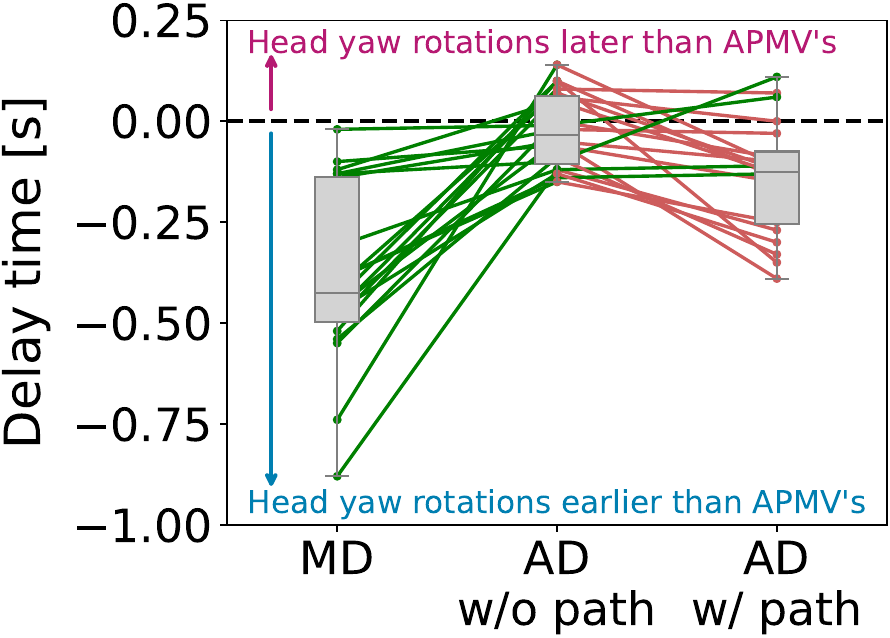}
    \caption{The delay time of irregular path group}
\end{subfigure}
\begin{subfigure}{0.45\textwidth}
    \centering
    \includegraphics[width=0.9\linewidth]{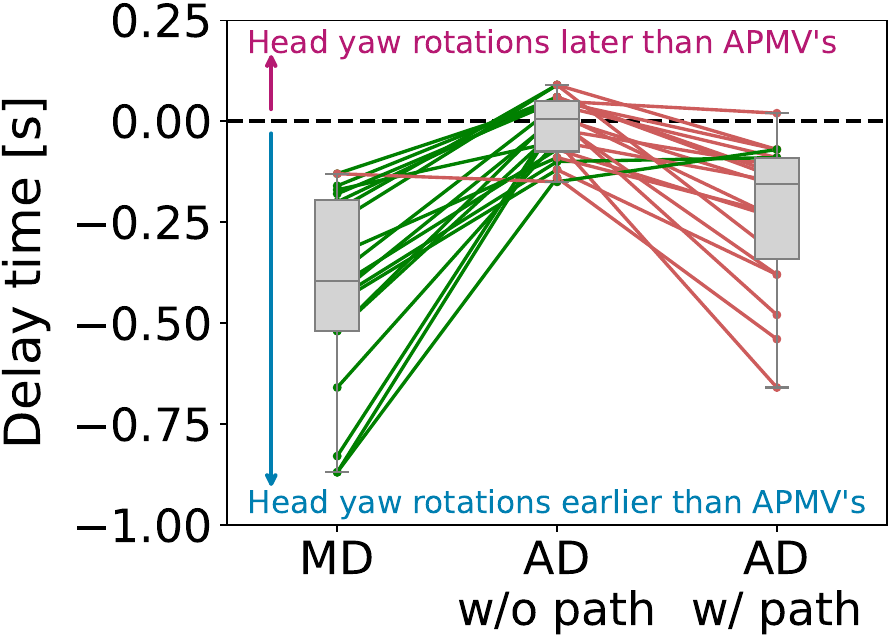}
    \caption{The delay time of regular path group}
\end{subfigure}
  \caption{Box-and-paired plots of the delay time across conditions. 
The box plots show the distribution of delay time for each condition
Color lines represent within-subject changes between adjacent conditions. In which, green lines indicate an increase from left to right, red lines a decrease, and gray lines no change.}
  \label{fig:Time_delay}

\centering
\captionof{table}{Results of the three-way ART ANOVA for the delay time based on a linear mixed-effects model}
\label{tab:ANOVA-Time_delay}
\begin{tabular}{lrccrr@{\hspace{2pt}}l}
\toprule
Dependent Variable & Source & dof & dof.res & \multicolumn{1}{c}{$F$} & \multicolumn{2}{c}{$p$} \\ \midrule
 \multirow{7}{*}{Delay time} 
 & Gender & 1 & 36 & 1.169 & .287 &  \\
 & Path & 1 & 36 & 0.211 & .649 &  \\
 & Condition & 2 & 72 & 95.226 & $< .001$ & $***$ \\
 & Gender $\times$ Path & 1 & 36 & 0.080 & .779 &  \\
 & Gender $\times$ Condition & 2 & 72 & 1.625 & .204 &  \\
 & Path $\times$ Condition & 2 & 72 & 1.321 & .273 &  \\
 & Gender $\times$ Path $\times$ Condition & 2 & 72 & 2.875 & .063 & \\
 \bottomrule
\multicolumn{7}{l}{\footnotesize $*$: $p<0.05$; $**$: $p<0.01$; $***$: $p<0.001$.} \\ 
\end{tabular}

\vspace{2mm}

\centering
\captionof{table}{ART-C Post-hoc comparisons with FDR adjustment for the main effect of driving conditions in the delay time}
\begin{tabular}{cccrrrrr@{\hspace{2pt}}l}
\toprule
Dependent Variable & Condition A & Condition B & estimate & SE & dof & t.ratio & \multicolumn{2}{c}{$p$-adj}  \\
\midrule
\multirow{3}{*}{Delay time}
 & AD w/ path   & AD w/o path & -35.8  & 4.58 & 72 & -7.812 & $< .001$ & $***$ \\
 & AD w/ path   & MD          &  27.2  & 4.58 & 72 &  5.946 & $< .001$ & $***$ \\
 & AD w/o path  & MD          &  63.0  & 4.58 & 72 & 13.758 & $< .001$ & $***$ \\
\bottomrule
\multicolumn{9}{l}{\footnotesize SE: standard error of the estimate.  $*$: $p<0.05$; $**$: $p<0.01$; $***$: $p<0.001$.}
\end{tabular}
\label{tab:posthoc-Time_delay}
\end{figure}

Fig.~\ref{fig:Time_delay} shows the delay time between the APMV yaw rotation and the participant's head yaw rotation across conditions for both the irregular and regular path groups.
Positive values indicate that head rotation occurred later than the APMV, whereas negative values indicate earlier head rotation relative to the APMV.
In both path groups, although the median delay time in the AD w/o path condition was close to zero, 
the third quartile was positive, indicating that at least approximately 25\% 
of the participants had head yaw rotations that occurred later than the APMV yaw rotations.
In contrast, in both of MD and AD w/ path conditions, the third quartile was negative, 
indicating that at least approximately 75\% of the participants 
had head yaw rotations that occurred earlier than the APMV yaw rotations.

The results of the three-way ART ANOVA for delay time are summarized in 
Table~\ref{tab:ANOVA-Time_delay}. 
A significant main effect of driving conditions was observed, 
whereas no significant main effects of genders or driving paths were found. 
No significant interaction effects were observed.
Post-hoc comparisons via ART-C for the main effect of driving conditions are presented in 
Table~\ref{tab:posthoc-Time_delay}. 
All pairwise comparisons were statistically significant.
This indicates that head yaw rotation in the MD condition occurred significantly earlier than in both AD conditions. 
Furthermore, head yaw rotation in the AD w/ path condition occurred significantly earlier than in the AD w/o path condition.

\clearpage

\subsection{Correlations between MISC Metrics and Delay time}

\begin{figure}[!h]
  \centering
    \begin{subfigure}[t]{0.45\linewidth}
    \centering
    \includegraphics[width=0.9\linewidth]{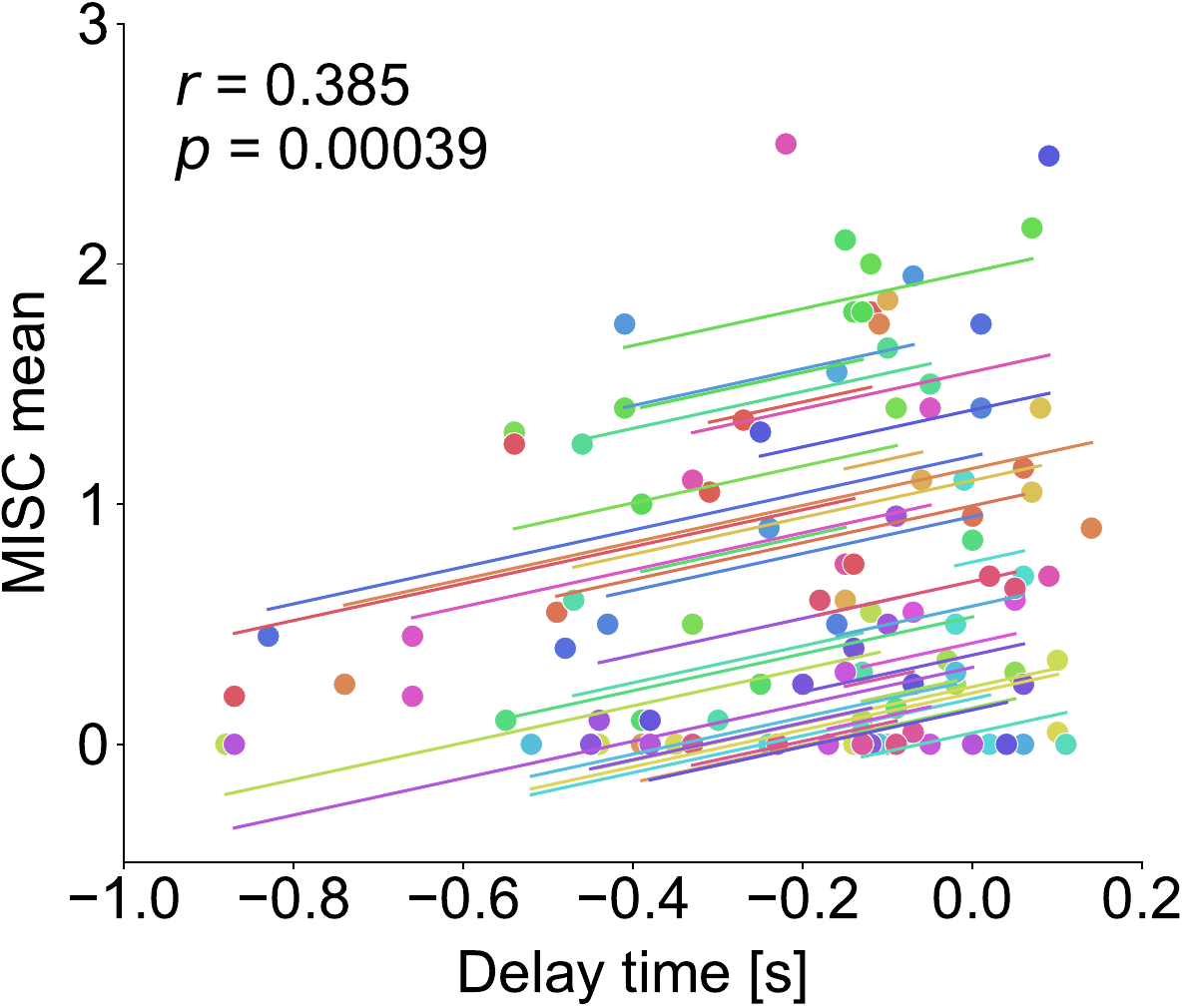}
    \caption{MISC mean and delay time}
  \end{subfigure}
\begin{subfigure}[t]{0.45\linewidth}
    \centering
    \includegraphics[width=0.9\linewidth]{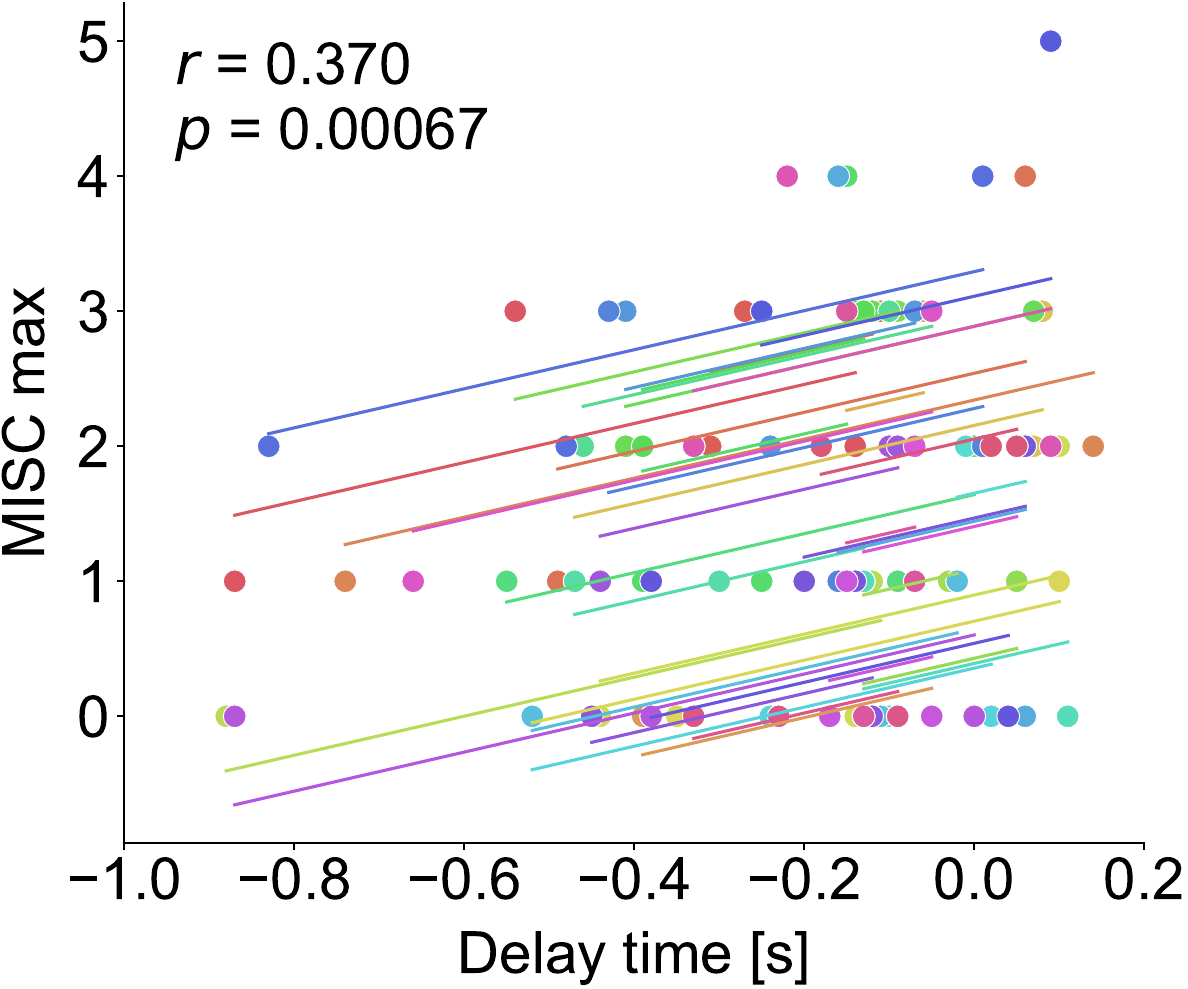}
    \caption{MISC max and delay time}
  \end{subfigure}
  \\
  \centering
    \begin{subfigure}[t]{0.45\linewidth}
    \centering
    \includegraphics[width=0.9\linewidth]{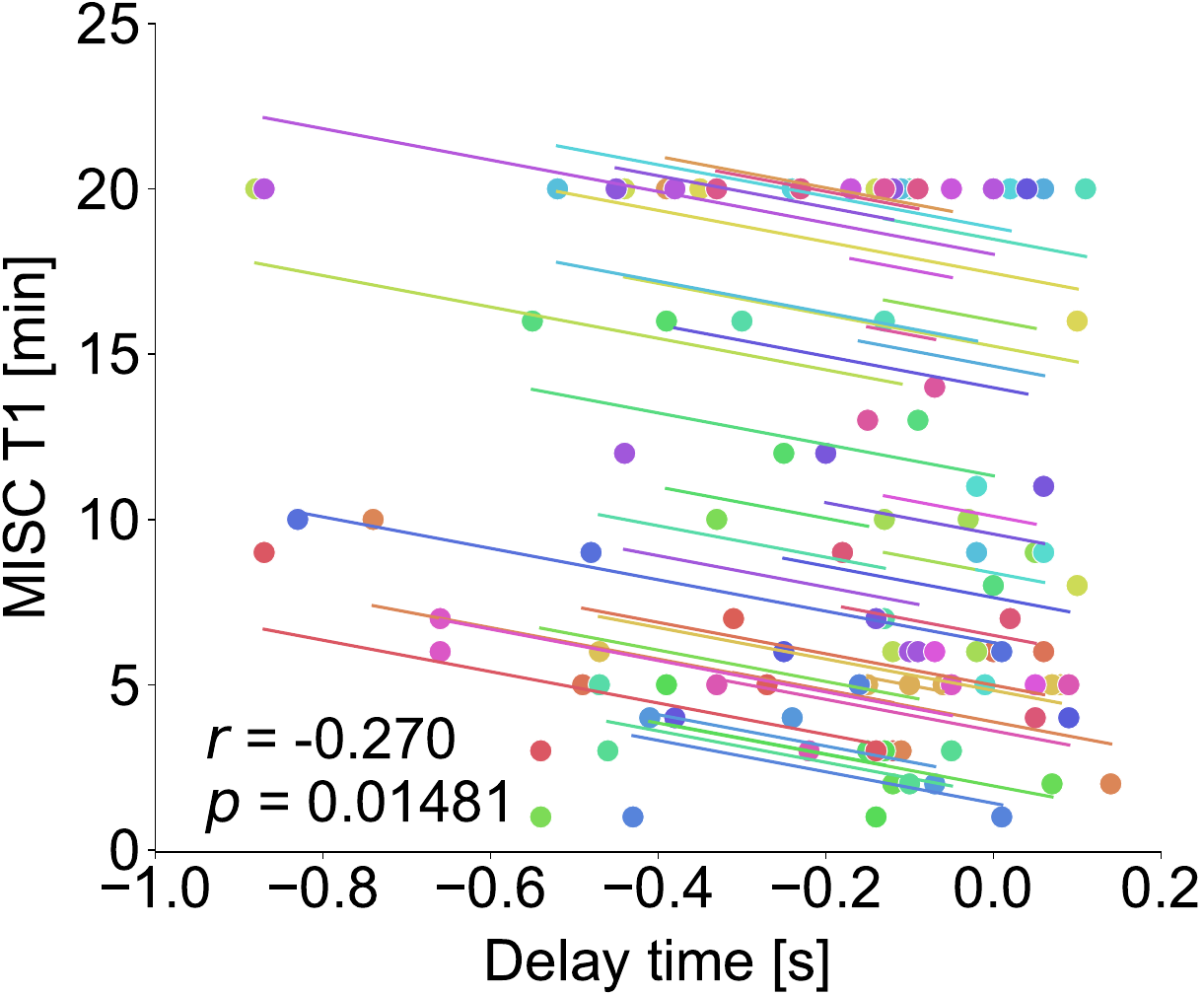}
    \caption{MISC T1 and delay time}
  \end{subfigure}
  \begin{subfigure}[t]{0.45\linewidth}
    \centering
    \includegraphics[width=0.9\linewidth]{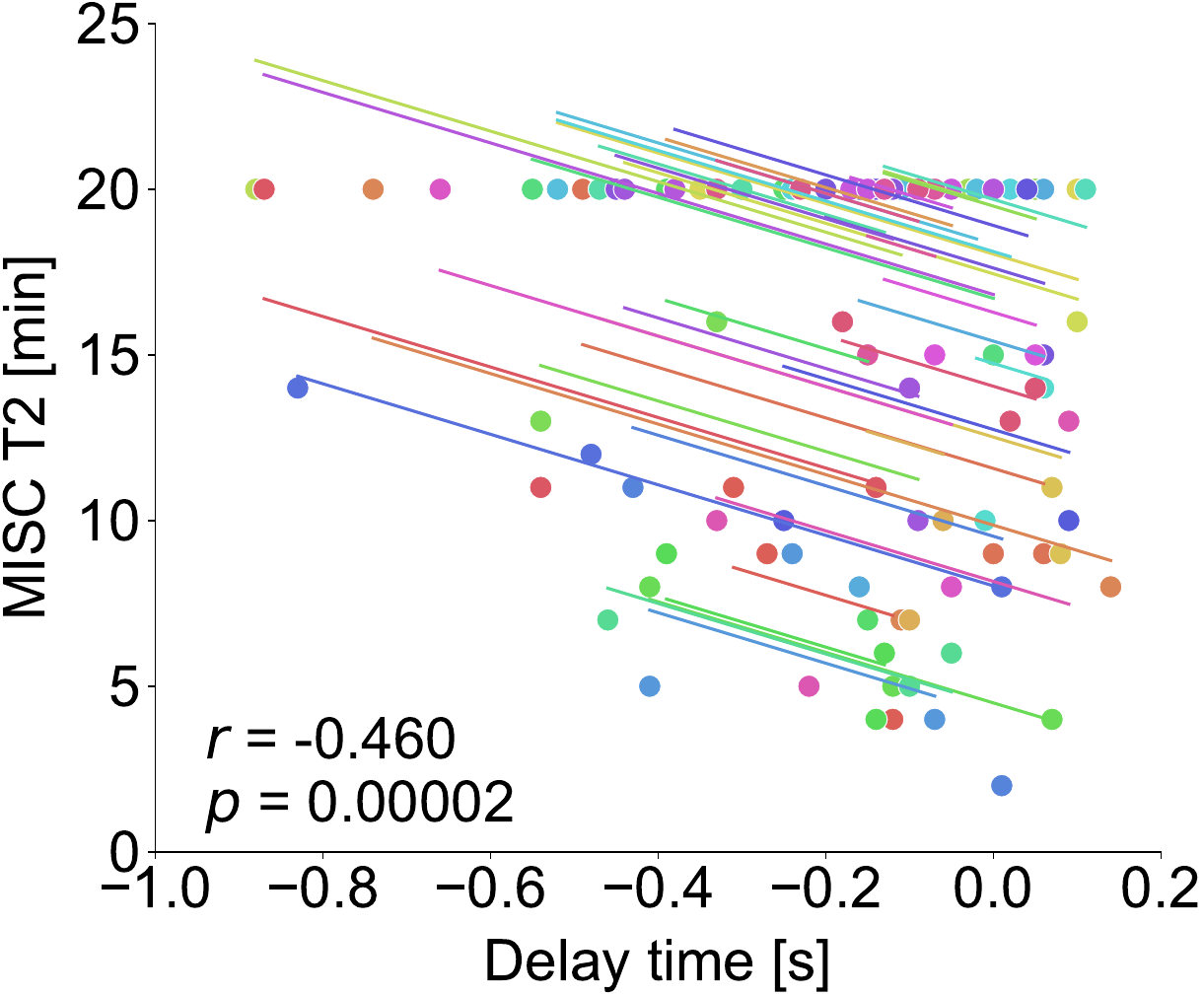}
    \caption{MISC T2 and delay time}
  \end{subfigure}
  \caption{Repeated-measures correlations between delay time and MISC metrics. Each color represents one participant, with dots indicating individual observations in three driving conditions and lines representing individual regression lines.}
  \label{fig:TimeDelay_MISCrmcorr}
\vspace{2mm}

\centering
\captionof{table}{Repeated-measures correlations between delay time and MISC metrics.}
\label{tab:rmcorr} 
\begin{tabular}{lrrr@{\hspace{2pt}}lclr}
\toprule
\multicolumn{1}{c}{Variable pair} & \multicolumn{1}{c}{r} & dof & \multicolumn{2}{c}{$p$} & 95\% CI & \textit{power} \\
\midrule
Delay time -- MISC mean & 0.385 & 79 &$<.001$ &***& [0.18, 0.56] & 0.950 \\
Delay time -- MISC max  & 0.370 & 79 & $<.001$ &***& [0.17, 0.54] & 0.932 \\
Delay time -- MISC T1   & -0.270 & 79 & $.015$ &*& [-0.46, -0.05] & 0.691 \\
Delay time -- MISC T2   & -0.460 & 79 & $<.001$ &***& [-0.62, -0.27] & 0.993 \\
\bottomrule
\multicolumn{7}{l}{\footnotesize CI: confidence interval.  $*$: $p<0.05$;  $***$: $p<0.001$.}
\end{tabular}
\end{figure}

To assess the association between the delay time and the four MISC metrics under all driving conditions, the results of the repeated-measures correlations are presented in Table~\ref{tab:rmcorr} 
and visualized in Fig.~\ref{fig:TimeDelay_MISCrmcorr}.
Repeated-measures correlation estimates a common slope across participants 
while allowing intercepts to vary. 
In the Fig.~\ref{fig:TimeDelay_MISCrmcorr}, 
each color represents one participant, 
with dots indicating individual observations under the three driving conditions 
and lines representing the common regression slope adjusted for each participant.
As shown in Table~\ref{tab:rmcorr}, delay time was significantly positively 
correlated with MISC mean and MISC max.
In contrast, significant negative correlations were observed between the delay time and both MISC T1 and MISC T2.

\section{DISCUSSION}

\subsection{Effect of Driving Path Indication on Motion Sickness and Head Motion}

Driving condition was the only factor that showed significant effects in the experiment. The three-way ART ANOVA revealed a significant main effect of driving condition across all MISC metrics (see Table~\ref{tab:ANOVA-MISC}) as well as on head motion delay time (see Table~\ref{tab:ANOVA-Time_delay}), with no significant interactions with gender or driving path. 
Specifically, both the MD and AD w/ path conditions showed lower motion sickness severity and longer motion sickness onset latency than the AD w/o path condition (see Table~\ref{tab:posthoc-MISC}). 
Similarly, the time delay between head motion and vehicle motion was significantly different among all three driving conditions, with participants' head motion occurring earliest relative to vehicle motion in the MD condition, followed by the AD w/ path condition, and latest in the AD w/o path condition (see Table~\ref{tab:posthoc-Time_delay}). 

Compared with the AD w/o path condition, participants in the MD condition reported significantly lower motion sickness severity, significantly longer motion sickness onset latency, and earlier head motion relative to vehicle motion during APMV turning. These two conditions may be interpreted as approximating the roles of a driver and a passenger, respectively. 
This pattern is consistent with previous studies showing that drivers are generally less susceptible to motion sickness than passengers \citep{rolnick1991driver,diels2016self,wada2018analysis}, as well as with studies suggesting that driver-like head movement strategies are associated with reduced motion sickness \citep{zikovitz1999head,wada2012can,wada2018analysis}.
In particular, previous studies have mainly discussed driver–passenger differences in terms of head movement direction, such as whether the head tilts toward the centripetal or centrifugal direction~\citep{zikovitz1999head,wada2012can,wada2018analysis}. 
In contrast, the present study highlights the temporal relationship between head motion and vehicle motion. 
A novel finding of this study is that participants in the MD condition initiated head yaw turns earlier than the yaw turns of the APMV, whereas in the AD w/o path condition their head yaw responses were significantly more delayed than those observed in the MD condition.
This suggests that the driver-like advantage in reducing motion sickness may be reflected not only in head movement direction, but also in temporally anticipatory head motions relative to vehicle motion, which may be associated with lower motion sickness severity and later symptom onset.

More importantly, participants in the AD w/ path condition also reported significantly lower motion sickness severity and significantly longer motion sickness onset latency comparing with the AD w/o path condition, with both measures reaching levels comparable to those in the MD condition (Table~\ref{tab:posthoc-MISC}). 
Moreover, when the driving path was indicated, participants showed significantly earlier head yaw turns relative to the APMV yaw turn than in the AD w/o path condition, although their head motion still lagged behind that in the MD condition (see Table~\ref{tab:posthoc-Time_delay}). 
These findings suggest that ground-based path indication during automated driving can effectively mitigate motion sickness in APMV passengers, while also shifting their head motion timing toward the driver-like pattern observed in MD. 
This interpretation is consistent with previous studies suggesting that improved anticipation of upcoming motion can alleviate motion sickness~\citep{diels2016self,ISKANDER2019716,kuiper2020knowing_unpredictable,wada2012can}. 
In particular, \citet{kuiper2020knowing_unpredictable} showed that passengers experience more severe motion sickness when motion is unpredictable in either its direction or the timing of directional changes.
Based on this point, the ground-based path indication proposed in the present study may have provided passengers with both types of anticipatory information. 
First, by displaying the future driving path on the ground, it conveyed the upcoming direction of vehicle motion.
Second, by showing where the vehicle trajectory would change, it may also have enabled passengers to infer the timing of the upcoming motion change by comparing the indicated path with the current vehicle motion ~\citep{griffin2004visual,tatsuno2024considerations}.
In this sense, the present path indication method may correspond well to the two aspects of anticipation highlighted by~\citep{kuiper2020knowing_unpredictable}, \ie anticipation of motion direction and anticipation of motion change timing.
This may explain why the AD w/ path condition reduced motion sickness and induced earlier head yaw responses than the AD w/o path condition.
Moreover, in terms of head motion, \citet{wada2012can} showed that passengers' motion sickness was significantly reduced when they actively imitated the driver-like strategy of tilting the head toward the centripetal direction.
This may partly explain why, in the present study, the AD w/ path condition was associated with both a more MD-like head motion and a significant reduction in motion sickness compared with the AD w/o path condition.

In summary, the present findings indicate that ground-based path indication can induce corresponding anticipatory head movements in passengers before changes in vehicle motion and can effectively mitigate motion sickness in APMV passengers during automated driving.


\subsection{Effects of Gender and Driving Path Pattern on Motion Sickness and Head Motion}

First, the present result of the three-way ART ANOVA results in Table~\ref{tab:ANOVA-MISC} showed no significant main effects or interactions involving gender across all MISC metrics. 
This finding is in line with previous experimental studies using controlled motion stimuli~\citep{cheung2003lack,klosterhalfen2006gender}, which similarly suggested that gender differences in motion sickness were not always evident.
However, this result appears to conflict with some previous findings. 
In contrast, a retrospective questionnaire-based study~\citep{lentz1977motion} reported greater motion sickness susceptibility in females than in males.
Similarly, \citet{golding2006motion} reviewed survey-based evidence suggesting that females are generally more susceptible to motion sickness than males, while also noting that the mechanisms underlying this sex difference remain unclear.
In line with these null effects on MISC, the ART ANOVA results for head motion delay time (see Table~\ref{tab:ANOVA-Time_delay}) also showed that there was no significant gender difference in the anticipatory timing of head motion relative to vehicle motion.


Second, although the regular path was expected to be inherently more predictable than the irregular path, no significant main effect of driving path or interactions involving driving path were observed across all MISC metrics (see Table~\ref{tab:ANOVA-MISC}). Consistent with the MISC results, no significant main effects or interactions involving driving path were observed for head motion delay time (see Table~\ref{tab:ANOVA-Time_delay}). 
This result differs from the findings of~\citep{henry2023car}, who reported that motion sickness in front-seat passengers increased with decreasing path predictability in a real-vehicle experiment comparing regular and unpredictable slalom paths.
One possible explanation for the discrepancy is that, although the irregular path in the present experiment contained various irregular turns, it formed a looped trajectory. As a result, over the course of the 15 minutes ride, the overall path pattern and vehicle's motion pattern may still have been relatively easy for participants to learn and anticipate.
However, participants' ability to recognize the path pattern and vehicle's motion pattern from riding experience was not explicitly assessed and remains to be clarified in future research.

\subsection{Exploratory Discussion: The Association Between Motion Sickness and Head Motion}

The above results showed that, across all experimental conditions involving two driving paths and three driving conditions, variations in participants’ MISC metrics tended to be accompanied by variations in head motion delay time. Accordingly, repeated-measures correlation analyses were conducted to examine the within-participant association between head motion delay time and each MISC metric.
Table~\ref{tab:rmcorr} and Fig.~\ref{fig:TimeDelay_MISCrmcorr} show that the delay time was significantly positively correlated with MISC mean and MISC max, and significantly negatively correlated with MISC T1 and MISC T2.
These results suggest that longer temporal delays between head motion and vehicle motion were associated with greater motion sickness severity and earlier symptom onset.
Interestingly, the correlation coefficient between MISC T2 and delay time was higher than that between MISC T1 and delay time, which may indicate that the delay time of head motion was more closely associated with the onset of more severe motion sickness symptoms than with the onset of mild symptoms.

Based on the above results, the present study may provide two possible insights into motion sickness reduction in autonomous mobility domain.
First, whereas previous studies have mainly emphasized head movement direction~\citep{zikovitz1999head,wada2012can,wada2018analysis}, the temporal relationship between head motion and vehicle motion may also contribute to motion sickness reduction.
Second, an earlier anticipatory head response relative to vehicle motion may be associated not only with reduced motion sickness severity but also with delayed symptom onset.

At present, although the association between motion sickness and head movements was observed in the present experiment, the underlying physiological mechanisms and the optimal timing of head motion anticipation relative to vehicle motion remain unclear and require further investigation.

\subsection{Limitations}

Because the participants in the present experiment were limited to individuals in their 20s and 30s, the age range was relatively narrow.
Accordingly, caution is needed in generalizing the present findings to other age groups, such as young children under 6 years of age and older adults over 60 years of age.

To prevent premature exposure to the APMV's driving path, the AD w/o path condition was always presented first to all participants. This design choice inevitably introduced potential order effects that may have influenced the results.

Additionally, it should be noted that the present study did not aim to investigate the design of an HMI for path presentation, but rather to examine the effect of driving path indication on motion sickness and head motion. 
Therefore, tape markings on the ground were used in the experiment to clearly present the path information. 
However, such an approach would be impractical for real-world APMV deployment.

Moreover, although the maximum speed and acceleration of the APMV were constrained to the same thresholds across all three conditions, the vehicle motion in the MD condition, in which participants manually drove the APMV, may not have been exactly the same as that in the two AD conditions. 
In addition, because the automated driving system mounted on the APMV could be affected by factors such as the initial state of the vehicle, passenger weight, and battery level, the vehicle motion during automated driving on the same path could still vary slightly across trials.
Although this setting may better reflect practical automated-driving scenarios, it also means that the motion stimuli experienced by participants were not completely identical even within the same condition.
This may have introduced slight variability in motion stimuli within the same condition and should therefore be considered a limitation of the present study.

Furthermore, although an association was observed in the present study between motion sickness and yaw-axis head motion in APMV occupants, the physiological mechanisms underlying this association, as well as any direct causal relationship, remain unclear.

\subsection{Future Work}
Future studies should explore more practical methods for presenting driving-path information to APMV occupants, such as the use of augmented reality displays or projection of the path onto the ground~\citep{zhang2024shared}.

Additionally, the optimal lead time for passengers to initiate head motion before vehicle turning remains unclear and should be examined in future studies.
Further optimization of this timing using a mathematically based motion sickness model, such as the 6-DoF SVC model~\citep{inoue2025construction}, will also be an important direction for future work.

\section{CONCLUSION}

This study investigated the effects of driving path indication on motion sickness and head motion in an APMV under two driving paths, \ie irregular and regular paths, and three driving conditions, \ie manual driving (MD), automated driving without path indication (AD w/o path), and automated driving with path indication (AD w/ path). 
The results showed that driving condition was the only factor that significantly affected both motion sickness and the delay time of head motion relative to vehicle motion. 
Compared with the AD w/o path condition, both the MD and AD w/ path conditions were associated with lower motion sickness severity, longer motion sickness onset latency, and earlier head motion relative to vehicle motion.
In addition, repeated-measures correlation further showed significant associations between the delay time of head motion and all MISC metrics.
These findings suggest that providing path information during automated driving may help mitigate motion sickness in APMV passengers, and that the temporal relation between head motion and vehicle motion may be an important characteristic associated with motion sickness in autonomous mobility.

\section*{Declaration of generative AI and AI-assisted technologies in the writing process}
During the preparation of this work, the author(s) used OpenAI ChatGPT (GPT-5.2) to proofread English texts.
After using this tool/service, the author(s) reviewed and edited the content as needed and take(s) full responsibility for the content of the publication.

\section*{CRediT author statement}

\textbf{Yuya~Ide}: Methodology, Software, Investigation, Validation, Formal Analysis, Visualization, Writing - Original Draft.
\textbf{Hailong~Liu}: Conceptualization, Methodology, Software, Investigation, Formal Analysis, Visualization, Supervision, Writing - Original Draft, Writing - review \& editing.
\textbf{Takahiro Wada}: Supervision,Methodology, Project administration, Writing - review \& editing.

\section*{Acknowledgment}
This work was supported by JSPS KAKENHI Grant Number 24H00298, Japan.

\bibliographystyle{cas-model2-names}
\bibliography{main.bib}           %
             
\end{document}